\documentclass[aps,pra,twocolumn,showpacs]{revtex4}
\usepackage{epsfig}
\usepackage{amsbsy,latexsym}
\usepackage{amsmath}
\usepackage{amssymb, mathrsfs}
\usepackage[mathscr]{eucal}
\usepackage{hyperref}
\usepackage{enumerate}
\usepackage[active]{srcltx}
\usepackage{xy}
\xyoption{matrix}
\xyoption{frame}
\xyoption{arrow}
\xyoption{arc}

\usepackage{ifpdf}
\ifpdf
\else
\PackageWarningNoLine{Qcircuit}{Qcircuit is loading in Postscript mode.  The Xy-pic options ps and dvips will be loaded.  If you wish to use other Postscript drivers for Xy-pic, you must modify the code in Qcircuit.tex}
\xyoption{ps}
\xyoption{dvips}
\fi

\entrymodifiers={!C\entrybox}

\newcommand{\bra}[1]{{\left\langle{#1}\right\vert}}
\newcommand{\ket}[1]{{\left\vert{#1}\right\rangle}}
\newcommand{\qw}[1][-1]{\ar @{-} [0,#1]}
\newcommand{\qwx}[1][-1]{\ar @{-} [#1,0]}
\newcommand{\cw}[1][-1]{\ar @{=} [0,#1]}

\newcommand{\gate}[1]{*+<.6em>{#1} \POS ="i","i"+UR;"i"+UL **\dir{-};"i"+DL **\dir{-};"i"+DR **\dir{-};"i"+UR **\dir{-},"i" \qw}

\newcommand{\measureD}[1]{*{\xy*+=<0em,.1em>{#1}="e";"e"+UR+<0em,.25em>;"e"+UL+<-.5em,.25em> **\dir{-};"e"+DL+<-.5em,-.25em> **\dir{-};"e"+DR+<0em,-.25em> **\dir{-};{"e"+UR+<0em,.25em>\ellipse^{}};"e"+C:,+(0,1)*{} \endxy} \qw}

\newcommand{\control}{*!<0em,.025em>-=-<.2em>{\bullet}}

\newcommand{\ctrl}[1]{\control \qwx[#1] \qw}

\newcommand{\targ}{*+<.02em,.02em>{\xy ="i","i"-<.39em,0em>;"i"+<.39em,0em> **\dir{-}, "i"-<0em,.39em>;"i"+<0em,.39em> **\dir{-},"i"*\xycircle<.4em>{} \endxy} \qw}

\newcommand{\multigate}[2]{*+<1em,.9em>{\hphantom{#2}} \POS [0,0]="i",[0,0].[#1,0]="e",!C *{#2},"e"+UR;"e"+UL **\dir{-};"e"+DL **\dir{-};"e"+DR **\dir{-};"e"+UR **\dir{-},"i" \qw}
\newcommand{\ghost}[1]{*+<1em,.9em>{\hphantom{#1}} \qw}

\newcommand{\lstick}[1]{*!R!<.5em,0em>=<0em>{#1}}
\newcommand{\ustick}[1]{*!D!<0em,-.5em>=<0em>{#1}}

\newcommand{\Qcircuit}{\xymatrix @*=<0em>}

\newcommand{\pureghost}[1]{*+<1em,.9em>{\hphantom{#1}}}

\def\d{\operatorname{d}}
\def\<{\langle}
\def\>{\rangle}

\def\Tr{\operatorname{Tr}}

\def\:{\hbox{\bf :}}

\def\dag{\dagger}
\def\geq{\geqslant}
\def\leq{\leqslant}
\def\map#1{\mathcal #1}
\def\sH{\mathcal{H}}

\def\Lin{\mathcal{L}}

\def\dim{\operatorname{dim}}

\def\qed{$\,\blacksquare$\par}
\def\kk{\rangle\!\rangle}
\def\bb{\langle\!\langle}

\newcommand{\one}{{I}}

\newcommand{\hilb}[1]{\mathcal{#1}}
\newcommand{\defset}[1]{{\sf #1}}

\newcommand{\ketbra}[2]{{\ket{#1}\bra{#2} }}
\newcommand{\KetBra}[2]{{\Ket{#1}\!\Bra{#2} }}
\newcommand{\BraKet}[2]{{\langle  \!  \langle {#1}\vert {#2} \rangle
    \! \rangle}}
\newcommand{\Bra}[1]{{ \langle \! \langle{#1}\vert }}
\newcommand{\Ket}[1]{{ \vert {#1}  \rangle \!  \rangle}}

\newtheorem{Def}{Definition}
\newtheorem{lemma}{Lemma}
\newtheorem{proposition}{Proposition}

\def\Proof{\medskip\par\noindent{\bf Proof. }}
\def\qed{$\blacksquare$}

\begin{document}
\title{Optimal processing of reversible quantum channels}
 \author{Alessandro Bisio}\email{alessandro.bisio@unipv.it}
 \affiliation{QUIT group, Dipartimento di Fisica, INFN Sezione di Pavia, via Bassi
   6, 27100 Pavia, Italy}
 \author{Giacomo Mauro D'Ariano}
 \affiliation{QUIT group,  Dipartimento di Fisica, INFN Sezione di Pavia, via Bassi
   6, 27100 Pavia, Italy}
 \author{Paolo Perinotti}
 \affiliation{QUIT group,  Dipartimento di Fisica, INFN Sezione di Pavia, via Bassi
   6, 27100 Pavia, Italy}
 \author{Michal Sedl\'ak}
 \affiliation{Department of Optics, Palack\'{y} University, 17. listopadu 1192/12, CZ-771 46 Olomouc, Czech Republic }
   \affiliation{Institute of Physics, Slovak Academy of Sciences, D\'ubravsk\'a cesta 9, 845 11 Bratislava, Slovakia}
 \date{ \today}
\begin{abstract}
  We consider the general problem of the optimal transformation of $N$
  uses of (possibly different) unitary channels to a single use of
  another unitary channel in any finite dimension. We show how the
  optimal transformation can be fully parallelized, consisting in a
  preprocessing channel followed by a parallel action of all the $N$
  unitaries and a final postprocessing channel.  Our techniques allow
  to achieve an exponential reduction in the number of the free
  parameters of the optimization problem making it amenable to an
  efficient numerical treatment. Finally, we apply our general results
  to find the analytical solution for special cases of interest like
  the cloning of qubit phase gates.

\end{abstract}

\maketitle
\section{Introduction}

In the past decades, considerable progress has been made in the
understanding of the mathemathical structure of quantum theory.
Recently the view of Quantum Theory as an operational probabilistic
theory \cite{ludwig} was revitalized by the success of quantum
information theory, which helped framing the operational
axiomatization program into an information theoretic context
\cite{hardy,fuchs,darcup,brassard}. This approach has been a fruitful
line of investigation \cite{masanes,bruknerdakic} and remarkably lead
to a derivation of Quantum theory from operational and informational
principles \cite{infoder}.  

The founding pillar of this view is the basic notion of test, that
includes as a special case that of preparation and observation. The
second ingredient defining an operational probabilistic theory is
provided by the rules for calculating the probabilities of the
experimental outcomes. In this perspective, transformations of quantum
states can be characterized by the minimal axioms that ensure
preservation of the probabilistic structure of quantum theory. Such
axioms require a transformation to be linear, trace non increasing and
completely positive, identifying possible events in a test with
\emph{quantum operations}, with \emph{quantum channels} representing
deterministic ones.

In quantum information applications, not only quantum states but also
transformations can often be considered as carriers of information,
e.g.~in the context of channel discrimination
\cite{acin,lopresti,sacchi,memeff,watrous}, gate programming
\cite{nielsen}, gate teleportation \cite{huelga,bartlett,huang},
process tomography \cite{weinstein,obrien,tomo,ziman} multi-round
quantum games \cite{gutoski}, standard quantum algorithms
\cite{deutsch,grover,shor}, as well as cryptographic protocols
\cite{yuen,dennis,pirandola}. This approach suggested to extend the
Kraus' axiomatic characterization of quantum operations to the case of
\emph{higher order quantum maps}, that is quantum maps that transform other
quantum maps. The easiest case of higher order quantum map 
is the \emph{supermap}, that is a map that transforms quantum
operations into quantum operations.
As a paradigmatic example, one can consider a supermap
that, given a single use of a quantum channel $\mathcal{T}$ as an
input, produces as an output channel $\mathcal T$ followed by a fixed
channel $\mathcal S$, namley $\mathcal S\circ\mathcal T$. It is
interesting to realize that \emph{higher order quantum computation},
namely the study of higher order quantum maps, is a subject in which
the differences between the quantum and the classical world are
evident. In classical computation, the Church-Turing paradigm of {\em
  program as data} allows one to compute functions of functions,
rather than only functions of bits. In the quantum case quantum data,
i.e. states, and quantum functions, i.e quantum transformations, are
intrinsically different objects and the exact programming of unitary
transformations via quantum states is impossible with finite
resources. Thus the study of the properties of higher order maps
achieves a twofold objective: on the one hand their mathemathical
characterization allows one to address in a systematic way all of the
quantum processing tasks, and on the other hand it provides new
insights in the distinctive features of quantum theory.

Higher order quantum maps were introduced in
Refs.~\cite{architecture,supermaps} and a complete axiomatic
characterization of a sub-hierarchy of the higher order quantum maps
was presented in Ref \cite{comblong}. Such a characterization is based
on the generalization of the notion of Choi operator to higher order
quantum maps. The subclass of higher order maps studied in
Ref.~\cite{comblong}, the so-called \emph{quantum combs}, was therein
proved to be in correspondence with the set of adaptive quantum
strategies, which are the most general architecture allowed in the
quantum circuit model. Such a unified description opened the way to
the formulation and optimization of a number of quantum processing
tasks \cite{cloning,tomo,learning,infotrade,clonmeas,learnmeas}.
However, there exist higher order maps which are admissible, i.e. they
preserve the probabilistic structure of quantum theory, but cannot be
described as a quantum circuit. For example, as pointed out in Ref.~
\cite{beyond} the map which receives one use of channel $\mathcal{C}$
and one use of channel $\mathcal{D}$ as input and outputs the convex
combination $\frac12 (\mathcal{C}\circ \mathcal{D} + \mathcal{D} \circ
\mathcal{C})$ is not realizable as a quantum circuit. This issue
raises two main questions. The first one, which is still completely
open, is which non-circuital higher order maps correspond to
physically feasible procedures. The second question asks whether there
exist any computational tasks in which this non-circuital higher order
map can outperform a circuital strategy. As regards this second
question it has indeed been proved that non-circuital maps can enhance
non-signalling gate discrimination \cite{giulio} and the
programmability of permutations of N different unitary channels
\cite{cdpf}.

Here we apply the theory of higher order quantum maps to the
optimization of a very general class of quantum information processing
tasks that can be sketched as follows. Let $\{
\mathcal{U}^{(i)}_g \}$ $g\in G$, $i=1,\ldots, N$, be a set of unitary
channels $\mathcal{U}^{(i)}_g (\rho)=U^{(i)}_g \rho \; U^{\dagger
  (i)}_g$, where $U^{(i)}_g$ is a unitary representation of a compact
group $G$ for each $i$.  Suppose that an unknown element $g \in G$ was
chosen randomly according to the Haar measure on $G$, and
conditionally on the outcome $g$ we had access to a single use of each
of the channels $\mathcal{U}^{(i)}_g$ $i=1,\ldots, N$, in any
sequential order. In other words, we can choose to use the channels
$\mathcal{U}^{(i)}_g$ in the sequence defined by any permutation
$\pi(i)$ of the indices $i$, and we are also free to use some of the
channels in parallel, in a single computational step. Our aim is now
to approximate as good as we can the channel $\mathcal{V}_g$ defined
by a different representation of $G$. In simple terms we are
considering a higher order map which transforms a single use of the
channels $\mathcal{U}^{(i)}_g$ into a single use of a channel
$\mathcal{V}_g$.  Quantum cloning of a unitary transformation is the
special case in which $\{ \mathcal{U}^{(i)}_g \} = \{ \mathcal{U}_g\}$
and $\mathcal{V}_g = \mathcal{U}_g \otimes \mathcal{U}_g$.  Since the
input consists of more than a single use of a channel, we should in
principle allow for non circuital maps, like the one that can exchange
the sequential order of the unitary channels.

In this paper
after a review of the main results in higher order
quantum computation in Section \ref{sec:highord}, we will
prove in Section \ref{parallelrules} that, surprisingly, the optimal
strategy for the class of tasks considered above is
realizable via a simple three steps procedure: i) application of a
fixed preprocessing channel $\mathcal{C}_1$, ii) parallel action of
the unknown channels $\mathcal{U}^{(i)}_g$ on some part of the output
of $\mathcal{C}_1$ and iii) action of a postprocessing channel
$\mathcal{C}_2$.  This means that there is no need for any kind of non
circuital quantum maps for the purpose of optimization of this kind of
task. Thanks to this result and to the symmetries of the problem in
Section \ref{sec:optcircuit} we will show how the optimization of the
circuit is reduced to the problem of finding the set of probability
distributions $p^a_K$, $\sum_K p^a_K=1$ maximizing the function
$\Phi(p_K^a)=\sum_K (\sum_a \sqrt{q_K^a p_K^a})^2$, where $q^a_K$ are
a set of coefficients determined by the problem that can be efficiently
calculated.  Once the parameters $p^a_K$ are found, a realization of the
optimal strategy can be found by the method of
Ref.~\cite{minimalalgo}.  The problem addressed in this paper is very
general and allows one to optimize wide variety of problems either
analytically or by simple numerical optimization. Some examples of
application of our results are presented in section
\ref{sec:examples}. Finally, section \ref{sec:conclusions} summarizes
our conclusions and possible future extensions of the work.

\section{Higher order quantum maps}\label{sec:highord}

A \emph{quantum supermap} is a transformation in which the input and
the output are quantum transformations themselves. In other words,
a higher order map describes a transformation that receives a
quantum operation as an input and produces another quantum operation
as an output, with the condition that channels are mapped to channels.
More generally one can consider maps whose input and output are
themselves supermaps, and the construction can be brought to
arbitrarily high order. In this way one obtains a whole hierarchy of
maps, the {\em higher order quantum maps}. In this section we review
the general theory of the higher order quantum maps, as presented in
Refs.~\cite{architecture, supermaps, comblong, actaphysica}, which we
refer to for an extensive discussion and for the proofs of the results
reviewed in this section.

The main issue addressed here is the classification of all the
input/output transformations that are admissible in principle
according to quantum theory. There are essentially two requirements
that an input/output map has to satisfy in order to be admissible.
The first one is linearity, which is required to be compatible with
the probabilistic structure of the theory. For example, let us
consider a supermap $\tilde{\mathcal{R}}$ which transforms channels
into channels, i.e. $\tilde{\mathcal{R}}: \mathcal{E} \mapsto
\tilde{\mathcal{R}}(\mathcal{E})$. If we apply the map
$\tilde{\mathcal{R}}$ to the convex combination $p \mathcal{E}_1 +
(1-p) \mathcal{E}_2$---corresponding to a random choice of the input
channel---the output has to be the convex combination $p
\tilde{\mathcal{R}}(\mathcal{E}_1) + (1-p)
\tilde{\mathcal{R}}(\mathcal{E}_2)$. For the same reason, we should
also have $\tilde{\mathcal{R}}(p \mathcal{E}) = p \tilde{\mathcal{R}}(
\mathcal{E})$ for any $0\leq p \leq 1$. These two conditions together
imply that $\tilde{\mathcal{R}}$ can be extended without loss of
generality to a linear map. The same reasoning used for supermaps
applies to more general higher order maps, which must then be linear
at every order. Actually, it is easy to show by induction
that every class of higher order quantum map corresponds to a convex
set. The second requirement is that the map must produce a
legitimate output when applied locally on one side of a bipartite
input. When the input and the output are quantum states this condition
is called complete positivity (CP) and the set of the admissible
maps is simply the set of the so called Quantum Operations \cite{Kraus}.

Let us now consider supermaps, whose input and output are quantum
operations. In order to simplify the presentation we will restrict
ourself to the deterministic case, that is maps  $\tilde{\mathcal{R}}$ which transform
quantum channels into quantum channels.
The generalization to the probabilistic case is possible and we refer
to
\cite{supermaps, comblong} for a comprehensive presentation.
If $\tilde{\mathcal{R}}$ is an admissible supermap transforming quantum channels with input
(output) space $\hilb{H}_{{\rm in}, A}$($\hilb{H}_{{\rm out}, A}$)
then the output of
$\tilde{\mathcal{R}}$ is a legitimate quantum channel even when
$\tilde{\mathcal{R}}$ is applied locally to a bipartite quantum channel,
i.e. a quantum channel $\mathcal{E}$ with bipartite input space
$\hilb{H}_{\rm in} := \hilb{H}_{{\rm in}, A}\otimes \hilb{H}_{{\rm
    in}, B}$ and bipartite output space $\hilb{H}_{\rm out} :=
\hilb{H}_{{\rm out}, A}\otimes \hilb{H}_{{\rm out}, B}$. This means that
$\tilde{\mathcal{R}}\otimes
\mathcal{I}_B(\mathcal{E})$ is a CP map for any bipartite quantum
channel $\mathcal{E}$, $ \mathcal{I}_B$ denoting the identity
map on the spaces labeled by $B$.

When dealing with complete positivity it is convenient to use the Choi
isomorphism \cite{choi} between $ \Lin(\Lin(\sH_{\rm in}),
\Lin(\sH_{\rm out}))$ and $\Lin(\sH_{out} \otimes \sH_{in})$, where
$\Lin(\sH)$ denotes the space of linear operators on the Hilbert space
$\sH$ and $\Lin(\Lin(\sH_{\rm in}), \Lin(\sH_{\rm out}))$ denotes the
space of linear maps from $\Lin(\sH_{\rm in})$ to $\Lin(\sH_{\rm
  out})$.  Before presenting the Choi isomorphism we recall the
following one to one correspondence between $\mathcal{L}(\hilb{H})$
and $\hilb{H} \otimes \hilb{H} $:
\begin{align}\label{eq:doubleket}
  &A = \sum_{nm}\bra{n}A\ket{m} \ketbra{n}{m}
\leftrightarrow
\Ket{A} = \sum_{nm}\bra{n}A\ket{m} \ket{n}\ket{m}& \nonumber\\
&A \otimes B \Ket{C} = \Ket{ACB^T},&
\end{align}
where $\ket{n}$ denotes a fixed orthonormal basis for $\hilb{H}$ and
$A^T$ denotes transposition of $A$ with respect to the fixed
orthonormal basis ($A^*$ denotes complex conjugation with respect to
the same basis).

\begin{proposition}[Choi isomorphism]
  Let $\mathfrak{C}$ be a linear map
from $\Lin(\Lin(\sH_{\rm in}), \Lin(\sH_{\rm out}))$
to $\Lin(\sH_{out} \otimes \sH_{in})$ defined as follows:
\begin{align}
  \label{eq:choiiso}
  \mathfrak{C}(\mathcal{C}) := \mathcal{ C} \otimes \mathcal{I}
  (\KetBra{I}{I}),
\end{align}
where $\Ket{I} \in \sH_{\rm in} \otimes \sH_{\rm in}$.
Then $\mathfrak{C}$ is invertible and its inverse map is defined as
\begin{align}
  \label{eq:invchoiiso}
  [ \mathfrak{C}^{-1}(C) ] (\rho) := \Tr_{\rm in}[(I_{\rm out} \otimes \rho^T)C ],
\end{align}
where $\Tr_{\rm in}$ denotes the partial trace over
$\hilb{H}_{\rm in}$ and $I_{\rm out}$ denotes the identity operator
over $\hilb{H}_{\rm out}$.
The operator $C := \mathfrak{C}(\mathcal{C})$ is called
the \emph{Choi operator} of the map $\mathcal{C}$.
\end{proposition}

For the special case of a unitary channel 
$\mathcal{Z} : \Lin(\hilb{H}_0) \to \Lin(\hilb{H}_1)$, 
$\mathcal{Z}(\rho) := Z \rho Z^\dagger$
Eq. \eqref{eq:choiiso} and Eq. \eqref{eq:doubleket} give
\begin{align}\label{eq:choiunit}
  \begin{split}
      \mathfrak{C}(\mathcal{Z}) &= \mathcal{Z} \otimes \mathcal{I}
  (\KetBra{I}{I}) =(Z \otimes I) \KetBra{I}{I} (Z^\dag \otimes I)= \\
&=\KetBra{Z}{Z} \qquad \Ket{Z} \in \hilb{H}_1 \otimes \hilb{H}_0.
  \end{split}
\end{align}

By means of the Choi isomorphism, for any map
$\tilde{\mathcal{R}}$ that transforms linear maps $\mathcal{E} : \Lin(\hilb{H}_1) \to \Lin(\hilb{H}_2)$
to linear maps
$\mathcal{E}' : \Lin(\hilb{H}_0) \to \Lin(\hilb{H}_3)$ we can introduce
the conjugate map $\mathcal{R}$
defined as follows:
\begin{align}\label{conjumap}
  \mathcal{R} := \mathfrak{C} \circ \tilde{\mathcal{R}} \circ  \mathfrak{C}^{-1},
\end{align}
that transforms the Choi operator $E$ of $\mathcal{E}$ into the Choi
operator $E'$ of $\mathcal{E}'$.  It is possible to show
\cite{supermaps} that the admissibility conditions for
$\tilde{\mathcal{R}}$ are equivalent to linearity and complete
positivity of $\mathcal{R} $. Moreover, since $\mathcal{R}$ is a
linear map from $\Lin(\hilb{H}_1 \otimes \hilb{H}_2)$ to
$\Lin(\hilb{H}_0 \otimes \hilb{H}_3)$ we can apply the Choi
isomorphism and introduce its Choi operator $R$.  For the sake of
simplicity we will systematically use the map $\mathcal{R}$ instead of
$\tilde{\mathcal{R}}$ and denote by $R$ the corresponding Choi
operator.  Within this framework we can give the following formal
definition of a higher order map.
\begin{Def}\label{axiocomb}
  A \emph{1-comb} on $(\hilb{H}_0, \hilb{H}_1) $ is the Choi operator
  of a linear CP map from $\Lin(\hilb{H}_0)$ to $\Lin(\hilb{H}_1)$. A
  {\em probabilistic 1-comb} is a $1$-comb corresponding to a quantum
  operation, and a {\em deterministic 1-comb} is a $1$-comb
  corresponding to a quantum channel. For $N \geq 2 $, a {\em
    $N$-comb} $R^{(N)}$ $(\hilb{H}_0,\dots, \hilb{H}_{2N-1}) $ is the
  Choi operator of an \emph{admissible $N$-map}, i.e. a linear
  completely positive map $\mathcal{R}^{(N)}$ that transforms
  $(N-1)$-combs on $(\hilb{H}_1,\dots, \hilb{H}_{2N-2})$ into
  $1$-combs on $(\hilb{H}_0, \hilb{H}_{2N-1}) $. A \emph{deterministic
    $N$-comb} is a $N$-comb corresponding to a map that transforms
  deterministic $(N-1)$-combs to deterministic $1$-combs. For $N,M
  \geq 1$ a \emph{$(N,M)$-comb} is the Choi operator of an
  \emph{admissible $(N,M)$-map}, i.e. a linear completely positive map
  $\mathcal{R}^{(N,M)}$ that transforms $N$-combs into $M$-combs.  A
  \emph{deterministic $(N,M)$-comb} is a $(N,M)$-comb corresponding to
  a $(N,M)$-map that transforms deterministic $N$-combs to
  deterministic $M$-combs.
An $(N,M)$-comb $S$ such that $S\leq \bar S$ for a deterministic $(N,M)$-comb $\bar S$ is called {\em probabilistic}.
\end{Def}
Notice that $N+1$-combs can be also denoted as $(N,1)$-combs.  By
recursively applying Def. \ref{axiocomb} one can define
\emph{admissible $(x,y)$-maps} where $x$ and $y$ are
previously defined map types, thus creating the whole hierarchy of
higher order maps. Also in this case deterministic and probabilistic 
$(x,y)$-combs can be straightforwardly defined.

In Def. \ref{axiocomb} we defined $N-$combs as operators $R^{(N)}$  acting on an ordered
sequence of Hilbert spaces $\bigotimes_{k=0}^{2N-1} \hilb{H}_k $.  Such a labeling
can be done by exploiting the following diagrammatic
representation of quantum combs
\begin{align}\label{sch:comb}
  \begin{aligned}
    \Qcircuit @C=1.3em @R=1.3em {
      &\ustick{0}&\multigate{1}{\;\;\;}&\ustick{1}\qw&&
 \ustick{2}  & \multigate{1}{\;\;\;} \qw&\ustick{3} \qw & &\dots &
 \ustick{2N-2}  &
 \multigate{1}{\;\;\;} \qw&\ustick{2N-1} \qw\\
      & &\pureghost{\;\;\;} &\qw &\qw &
\qw &\ghost{\;\;\;}& \qw & &\dots &
 &
\ghost{\;\;\;}&}
  \end{aligned}
\end{align}
where an $N$-comb is represented by a
comb-like diagram with $N$ teeth.

The following proposition provides an algebraic characterization of the set
of deterministic $N-$combs.
\begin{proposition}\label{prop:normcomb}
A positive operator $R^{(N)}$ on $ \bigotimes_{k=0}^{2N-1} \hilb{H}_k $ is a
deterministic $N$-comb if and only if the following conditions hold:
\begin{align}\label{Eq:normcomb}
  \begin{split}
&\Tr_{2j-1}[R^{(j)}] = I_{2j-2} \otimes R^{(j-1)}, \qquad 2 \leq j
 \leq N\\
 &\Tr_1[R^{(1)}] = I_0,
  \end{split}
\end{align}
where $ R^{(j-1)}$, $2 \leq j \leq N$ are deterministic $(j-1)$-combs.
\end{proposition}
Proposition \ref{prop:normcomb} characterizes the set of deterministic
$N$-combs as the set of positive operators subject to the linear
constraints of Eq. \eqref{Eq:normcomb}. This implies that the set of
deterministic $N$-combs is a convex set.  It is possible to provide a
generalization of proposition \ref{prop:normcomb} to $(N,M)$-maps and
to all the other classes of higher order maps, but this is beyond the
main scope of this paper and we will omit it. However, let us
remind that each set of deterministic higher order maps is a convex
set.

So far we focused our analysis on the mathematical description of the
higher order quantum maps which culminated in Proposition
\ref{prop:normcomb}, which translates the admissibility conditions of
linearity and complete positivity in terms of algebraic constraints.
However, such a characterization would be just an abstract and rather
sterile construction if it was not related to physical achievability of
the involved maps.  In the following we will show that any admissible
deterministic $N$-map has a physical realization as a concatenation of
channels with multipartite input and output.

When considering channels whose input
and output spaces are tensor products of Hilbert spaces it is possible
to define the composition of these channels only through some of these
spaces.
For example, if we have $\mathcal{E} \in
\mathcal{L}(\mathcal{L}(\hilb{H}_0\otimes \hilb{H}_2),\mathcal{L}(\hilb{H}_1\otimes \hilb{H}_3))$
and
$\mathcal{D} \in
\mathcal{L}(\mathcal{L}(\hilb{H}_3\otimes
\hilb{H}_5),
\mathcal{L}(\hilb{H}_4\otimes \hilb{H}_6))$
it is possible to define the composition
\begin{align}
  \mathcal{D} \star \mathcal{E} :=
(\mathcal{D}\otimes \mathcal{I}_1) \circ (\mathcal{E} \otimes
\mathcal{I}_5)
 \label{eq:compositionoftwomap},
\end{align}
where $\mathcal{D} \star \mathcal{E}\in  \mathcal{L}(\mathcal{L}(\hilb{H}_0\otimes \hilb{H}_2\otimes \hilb{H}_5),\mathcal{L}(\hilb{H}_1\otimes \hilb{H}_4\otimes \hilb{H}_6))$. 
It can be diagrammatically represented as follows:
\begin{align}
  \begin{aligned}
    \Qcircuit @C=2em @R=1.5em {
      \ustick{0} & \multigate{1}{\mathcal{E}} & \ustick{1} \qw  & \ustick{5} &  \multigate{1}{\mathcal{D}} & \ustick{4} \qw   \\
      \ustick{2} & \ghost{\mathcal{E}} & \ustick{3} \qw &  \qw &\ghost{\mathcal{D}} & \ustick{6} \qw\\
    }
  \end{aligned}\quad.
\end{align}
Moreover, here the similarity with Eq. \eqref{sch:comb} is not a
coincidence as it will be clear later.  Since the two channels can be
represented in terms of their Choi operators one can reasonably wonder
how the Choi operator of the composition $ \mathcal{D} \star
\mathcal{E}$ can be expressed in terms of the Choi operators $D$ and
$E$.  For this purpose it is convenient to define the following
operation.

\begin{Def}\label{def:link}
  Let $M$ be an operator in $\mathcal{L}(\bigotimes_{i\in \defset{I}}\hilb{H}_i)$
and $N$ be an operator in $\mathcal{L}(\bigotimes_{j\in \defset{J}}\hilb{H}_j)$
where $\defset{I}$ and $\defset{J}$ are two finite sets of indexes.
Then the \emph{link product} $M*N$ is an operator in
$\mathcal{L}(\hilb{H}_{\defset{I}\setminus \defset{J}}\otimes \hilb{H}_{\defset{J}\setminus \defset{I}} )$
defined as
\begin{align}
  M*N := \Tr_{\defset{I} \cap \defset{J}}[(I_{\defset{J}\setminus \defset{I}} \otimes M^{T_{\defset{I} \cap \defset{J}}})
(I_{\defset{I}\setminus \defset{J}} \otimes N) ]
\label{eq:link}
\end{align}
where $\defset{A}\setminus \defset{B}:= \{i \in \defset{A} | i \notin  \defset{B}  \}$
and we introduced the notation
$\hilb{H}_{\defset{A}} := \bigotimes_{i\in \defset{A}}\hilb{H}_i$
for any set of indexes $\defset{A}$.
\end{Def}

It is worth noting that the link product is commutative,
i.e. $M*N = N*M$ (here we assume the same ordering of the tensor products of Hilbert spaces).
Moreover, the special case $\defset{I} \cap \defset{J} = \emptyset$
gives $N*M = N \otimes M$ while if $\defset{I} = \defset{J}$
$N*M = \Tr[M^TN]$.
The use of the link product simplifies the expression for the Choi operator of
the composition of two channels, which is the content of the following
Lemma.

\begin{lemma} \label{lem:compolink}
Let $\defset{in}_{\mathcal{E}}, \defset{out}_{\mathcal{E}}, \defset{in}_{\mathcal{D}}, \defset{out}_{\mathcal{D}}$
be four sets of indexes such that $\defset{in}_{\mathcal{E}} \cap \;
\defset{out}_{\mathcal{D}} = \emptyset$.
Let $\mathcal{E} \in \mathcal{L}(\mathcal{L}(\hilb{H}_{\defset{in}_{\mathcal{E}}})),
\mathcal{L}(\hilb{H}_{\defset{out}_{\mathcal{E}}})$
and $\mathcal{D} \in \mathcal{L}(\mathcal{L}(\hilb{H}_{\defset{in}_{\mathcal{D}}})),
\mathcal{L}(\hilb{H}_{\defset{out}_{\mathcal{D}}})$
be a couple of quantum channels.
Let $E$ and $D$ be Choi operators of the quantum channels
$\mathcal{E}$ and $\mathcal{D}$.
Then the Choi operator of the composition
\begin{align}\label{eq:choiofcomp}
\mathcal{D} \star \mathcal{E} :=
(\mathcal{I}_{\defset{out}_{\mathcal{E}}\setminus \defset{in}_{\mathcal{D}}}
\otimes
\mathcal{D})
\circ
(\mathcal{I}_{\defset{in}_{\mathcal{D}}\setminus \defset{out}_{\mathcal{E}} }
\otimes
\mathcal{E})
\end{align}
is given by
\begin{align}
\label{eq:choiofcomp1}
\mathfrak{C}(\mathcal{D} \star \mathcal{E}) = D*E
\end{align}
where $D*E$ is the link product of the two operators.
\end{lemma}

For sake of clarity, it is useful to apply Lemma
\ref{lem:compolink} to the simple case of two unitary channels
\begin{align}
  \begin{aligned}
    \Qcircuit @C=1.5em @R=1.5em {
      \ustick{1} & \gate{\mathcal{U}} & \ustick{2} \qw  &  \gate{\mathcal{V}} & \ustick{3} \qw
    }
  \end{aligned}\quad.
\end{align}
where
$\mathcal{U} \in \Lin(\Lin{(\hilb{H}_1}), \Lin{(\hilb{H}_2}))$
and $\mathcal{V} \in \Lin(\Lin{(\hilb{H}_2)}, \Lin{(\hilb{H}_3)})$.
Reminding Eq. \eqref{eq:choiunit} the Choi operators of
$\mathcal{U}$
and $\mathcal{V}$ are given by
$\KetBra{U}{U} \in \Lin(\hilb{H}_2\otimes\hilb{H}_1)$ and $\KetBra{V}{V} \in \Lin(\hilb{H}_3\otimes\hilb{H}_2)$,
respectively. By applying Eq. \eqref{eq:choiofcomp1}
we have
\begin{align}
  \begin{split} \label{eq:connectedunit}
&\mathfrak{C}(\mathcal{U} \star \mathcal{V} ) =
\KetBra{U}{U} *\KetBra{V}{V} =\\
&=\Tr_2[ (\KetBra{U}{U} \otimes I_3)  (I_1 \otimes (\KetBra{V}{V})^{T_2})] \\
&=
\Tr_2[ (\KetBra{U}{U} \otimes I_3)  (I_1 \otimes \KetBra{V^*}{V^*})]=\\
&=
\KetBra{UV}{UV}=\mathfrak{C}(\mathcal{U} \circ \mathcal{V} )
  \end{split}
\end{align}
where we used Eq. \eqref{eq:choiiso}.

Lemma \ref{lem:compolink} can be 
applied to the case in which
$N$ quantum channels are connected in a sequence, i.e
\begin{align}\label{eq:networkascircuit}
&  \mathcal{R} = \mathcal{E}_1 \star \mathcal{E}_2 \star \cdots \star \mathcal{E}_N
\end{align}
\begin{align}
&\begin{aligned}
\Qcircuit @C=1.2em @R=1em {
\ustick{0}&\multigate{1}{\mathcal{E}_1}& \ustick{1} \qw & \ustick{2}&
 \multigate{1}{\mathcal{E}_2} &  \ustick{3} \qw \\
& \pureghost{\mathcal{E}_1}& \ustick{A_1} \qw&\qw& \ghost{\mathcal{E}_1}  &
\ustick{A_2}\qw\\
}
\end{aligned}
\qquad\cdots\qquad
\begin{aligned}
\Qcircuit @C=1.2em @R=1em {
\ustick{2N-2}&   \multigate{1}{\mathcal{E}_N} &  \ustick{2N-1} \qw \\
 \ustick{A_{N-1}} & \ghost{\mathcal{E}_N}  &
\\
}
\end{aligned}
\nonumber
\end{align}
where $\mathcal{E}_i: \mathcal{L}(\hilb{H}_{2i-2}\otimes \hilb{H}_{A_{i-1}})
\rightarrow \mathcal{L}(\hilb{H}_{2i-1}\otimes \hilb{H}_{A_{i}})$,
$\hilb{H}_{A_0} = \hilb{H}_{A_N} = \mathbb{C}$ and the
ordering in which the connections are performed can be proved to be irrelevant.
Moreover the Choi operator of the sequence
$ \mathcal{R} = \mathcal{E}_1 \star \cdots \star \mathcal{E}_N$
 becomes
 \begin{align}
   \label{eq:multilink}
   \mathfrak{C}(\mathcal{R}) := R = E_1 * \cdots * E_N
 \end{align}
and also in this case the order in which the link
 products are performed is not relevant.
It is possible to prove \cite{comblong} that
Eq. \eqref{eq:multilink} implies that
the Choi operator of a sequence of channels
satisfies conditions \eqref{Eq:normcomb}. Moreover,
Eq. \eqref{Eq:normcomb} is a sufficient condition
for $R^{(N)}$ to be the Choi operator of a sequence of quantum
channels.
It is then possible to identify the set of admissible deterministic
$N-$maps with the set of maps that are given by the concatenation of
$N$ channels.

\begin{proposition}\label{prop:realization}
Let $\mathcal{R}^{(N)}$ be a linear map
and $R^{(N)}$ its Choi operator.
  Then the following conditions are equivalent:
  \begin{itemize}
  \item $R^{(N)}$ is a deterministic $N$-comb,
  \item there exist $N$ quantum channels
$\mathcal{E}_i: \mathcal{L}(\hilb{H}_{2i-2}\otimes \hilb{H}_{A_{i-1}})
\rightarrow \mathcal{L}(\hilb{H}_{2i-1}\otimes \hilb{H}_{A_{i}})$,
$(\hilb{H}_{A_0} = \hilb{H}_{A_N} = \mathbb{C})$, $i = 1, \dots,N$
such that
$ \mathcal{R}^{(N)} = \mathcal{E}_1 \star \cdots \star \mathcal{E}_N$.
 \end{itemize}
Moreover, for any deterministic $(N-1)-$comb
$T^{(N-1)}$ the transformation
\begin{align*}
\mathcal{R}^{(N)}:\mathcal{T}^{(N-1)} \mapsto
\mathcal{T}'^{(1)} := \mathcal{R}^{(N)}(\mathcal{T}^{(N-1)})
\end{align*}
is achieved by connecting the two sequences of channels as follows
\begin{align}\label{eq:linkmaps}
\begin{split}
&\mathcal{R}^{(N)}(\mathcal{T}^{(N-1)}) :=
\mathcal{R}^{(N)} \star \mathcal{T}^{(N-1)} = \\
\\
&=  \begin{array}{c}
  \mathcal{T}^{(N-1)} \\
\overbrace{\underbrace{
  \begin{aligned}
\Qcircuit @C=0.4em @R=0.5em {
&&  \multigate{1}{\mathcal{D}_1}& \qw &   \qw \\
&\multigate{1}{\mathcal{E}_1}& \ghost{\mathcal{D}_2} &  \multigate{1}{\mathcal{E}_2} &   \qw \\
& \pureghost{\mathcal{E}_1}&  \qw& \ghost{\mathcal{E}_1}  &
\qw\\
}
\end{aligned}
\cdots
\begin{aligned}
\Qcircuit @C=0.4em @R=0.5em {
&\multigate{1}{\mathcal{D}_{N-1}}&\qw &      \\
&\ghost{\mathcal{D}_{N-1}}&   \multigate{1}{\mathcal{E}_N}&\qw    \\
&& \ghost{\mathcal{E}_N}&
\\
}
\end{aligned}
}}\\
 \mathcal{R}^{(N)}
  \end{array}
\end{split}
\end{align}
and the Choi operator of the resulting map is given by:
\begin{align}
  \label{eq:linkchoi}
\mathfrak{C}(\mathcal{R}^{(N)} \star \mathcal{T}^{(N-1)}) =
R^{(N)} * T^{(N-1)}.
\end{align}
\end{proposition}

Proposition \ref{prop:realization} shows that any admissible deterministic $N$-map
has a physical realization as a concatenation of quantum
channels and tells us through Eq. \eqref{eq:linkchoi} that the action
of an admissible $N$-map on a
$N-1$-map can be expressed by the link product of the corresponding
Choi operators.

Unfortunately, the more general case of $(N,M)$-maps or $(x,y)$-maps
is more involved.  Eq. \eqref{eq:linkchoi} still holds, but it is no
longer possible to interpret $(N,M)$-maps or $(x,y)$-maps as sequences
of channels.

The following lemma can be regarded as
a quantum generalization of the
uncurrying procedure of the functional calculus and provides some useful insight
on the features of the deterministic $(N,M)$ maps.

\begin{lemma}\label{lem:uncurrying}
  Let $\mathcal{R}^{(N,M+1)}$ be an admissible deterministic
  $(N,M+1)$-map.  Then $\mathcal{R}^{(N,M+1)}$ is in one-to-one
  correspondence with an admissible map $\mathcal{R}^{(N \otimes M,1)}$ that transforms tensor
  product operators $S^{(N)} \otimes T^{(M)}$ of deterministic $N$
  and $M$-combs
 into deterministic $1$-combs.
\end{lemma}

Intuitively, the tensor product comb $S^{(N)}\otimes T^{(M)}$ can be
seen as couple of combs, one with $N$ teeth and the other with $M$
teeths, which create a $N+M$-comb where the order between two teeth of
different comb is not completely fixed, but only restricted by the two
orderings of the combs $S^{(N)}$ and $T^{(M)}$. Here follows a
pictorial example
\begin{align*}
S^{(2)}\otimes T^{(2)} =  \begin{aligned}
    \Qcircuit @C=0.4em @R=0.7em {
      &                        &               &     &
     \pureghost{\;\!\!\!}             &  \qw   &\qw &\qw                  &
\qw& \qw& \ghost{\;\!\!\!}\\
     & \multigate{1}{\;\!\!\!}& \qw         &      &
     \multigate{-1}{\;\!\!\!} &  \qw &       & \multigate{1}{\;\!\!\!} &
\qw&       & \multigate{-1}{\;\!\!\!}&\qw\\
     & \pureghost{\;\!\!\!}         & \qw &\qw&
     \qw                     &  \qw &\qw & \ghost{\;\!\!\!}           &
      &       &
}
  \end{aligned}\;\mbox{or}\;
  \begin{aligned}
    \Qcircuit @C=0.4em @R=0.7em {
      &                        &               &     &
     \pureghost{\;\!\!\!}             &  \qw   &\qw &\ghost{\;\!\!\!}               &
& & \\
     & \multigate{1}{\;\!\!\!}& \qw         &      &
     \multigate{-1}{\;\!\!\!} &  \qw &       & \multigate{-1}{\;\!\!\!} &
\qw&       & \multigate{1}{\;\!\!\!}&\qw\\
     & \pureghost{\;\!\!\!}         & \qw &\qw&
     \qw                     &  \qw &\qw &        \qw   &
      \qw&     \qw  &\ghost{\;\!\!\!}
}
  \end{aligned}
\; \dots
\end{align*}
This feature can be rephrased by saying that the tensor product comb
$S^{(N)}\otimes T^{(M)}$ is not endowed with a full definite causal
order.  An admissible map $\mathcal{R}^{(N\otimes M,1)}$ can in
principle exploit this freedom and convex combination or quantum
superposition of different causal orderings are allowed, like e.g.
\begin{align}
C\otimes D=
\begin{aligned}
    \Qcircuit @C=0.4em @R=0.7em {
&\gate{\mathcal{C}}& \qw &&&
&\gate{\mathcal{D}}& \qw &
}
  \end{aligned}\;\mbox{or}\;
  \begin{aligned}
    \Qcircuit @C=0.4em @R=0.7em {
&\gate{\mathcal{D}}& \qw &&&
&\gate{\mathcal{C}}& \qw &
}
  \end{aligned} \nonumber \\
R^{(1 \otimes 1,1)} * (C\otimes D)=
\frac12
  \begin{aligned}
    \Qcircuit @C=0.4em @R=0.7em {
&\gate{\mathcal{D}}&\gate{\mathcal{C}}& \qw &
}
  \end{aligned} + \frac12
  \begin{aligned}
    \Qcircuit @C=0.4em @R=0.7em {
&\gate{\mathcal{C}}&\gate{\mathcal{D}}& \qw &
}\; .
  \end{aligned}
\label{eq:cswitch}
\end{align}

It is possible to prove that the admissible $(1\otimes 1,1)$-map
defined by Eq. \eqref{eq:cswitch}, cannot be realized as a
concatenation of channels.  In Ref.~\cite{beyond} the first example of
an admissible deterministic $(N,M)$-map that cannot be realized as a
sequence of channels, has been found.  Even if a $(N,M)$-map
$R^{(N,M)}$ does not correspond to a sequence of channels this does
not imply that $R^{(N,M)}$ is not physically realizable. The
$(1\otimes 1,1)$-map in Eq. \eqref{eq:cswitch} receives in input one
use of channel $\mathcal{C}$ and one use of channel $\mathcal{D}$ and
outputs either $\mathcal{C} \circ \mathcal{D} $, or $\mathcal{D} \circ
\mathcal{C} $ with probability $\frac12$: this is clearly a well
defined operational procedure.  The characterization of admissible
$(N,M)$-maps that are not a sequence of channels, but nevertheless are
physically realizable, is still an open problem.  Recently
$(N,M)$-maps have been studied in Refs.~\cite{cdpf,giulio} where it
was shown that $(N,M)$-maps can enhance information processing tasks
like controlled permutation of oracle gates or discrimination between
no-signalling channels. Also the analysis of quantum correlations
without a pre-defined causal order in Ref.~\cite{brukner} can be
appropriately phrased in terms of $(N,M)$-maps.

\section{Processing of unitary transformations}

An example of a task which one can naturally address in the framework
of the higher order quantum maps, is cloning of a transformation.
This problem was for the first time introduced in Ref.~\cite{cloning}
and can be illustrated as follows.  Consider a user who is provided
with a single use of an unknown transformation $\mathcal{T}$. Suppose
now that he needs to run $\mathcal{T}$ twice in order to accomplish
some desired computational task. Then it would be extremely valuable
for him to have a ``cloner of transformations'' available. Such a
cloner would be a machine which receives a single use of the
transformation $\mathcal{T}$ as an input and outputs two copies of the
same transformation, i.e. $\mathcal{T} \otimes \mathcal{T}$.  In
Ref.~\cite{cloning} a no-cloning theorem for transformations was
proved and the optimal cloning map for the case in which the unknown
transformation is a unitary in $\rm{SU}(d)$ was derived. The optimal
cloner is an admissible deterministic $2$-map (see Def.
\ref{axiocomb}) which thanks to Proposition \ref{prop:realization} is
realizable as concatenation of channels.

In this section we consider a more general scenario which nevertheless
is closely related to the cloning of a unitary transformation.
Suppose that a user can have access to $N$ unknown unitary channels
$\{ \mathcal{U}^{(i)}_g \}_{i=1 \dots N}$, where $ \mathcal{U}^{(i)}_g
\in \Lin(\Lin(\hilb{H}_{2i-1}),\Lin(\hilb{H}_{2i}))$ and we denote by
$d_k$ the dimension of $\hilb{H}_k$.  The action of
$\mathcal{U}^{(i)}_g$ on a state $\rho $ is described by a unitary
representation ${U}^{(i)}_g $ of a fixed compact group $G$, i.e.
$\mathcal{U}^{(i)}_g(\rho) = {U}^{(i)}_g \rho {U}^{(i) \dag}_g$.

The task is to exploit the uses of the unitary channels $\{
\mathcal{U}^{(i)}_g \}_{i=1 \dots N}$ to create a target unitary
channel $\mathcal{V}_g : \Lin(\hilb{H}_0) \to \Lin(\hilb{H}_{2N+1})$
which is described by a different unitary representation $V_g$ of the
same group $G$.  The special case in which $\mathcal{U}^{(i)}_g =
\mathcal{U}_g$ $\forall i = 1,\dots,N$, and $\mathcal{V}_g =
\mathcal{U}_g^{\otimes M}$ corresponds to the cloning of a unitary
transformation $\mathcal{U}_g$ from $N$ copies to $M$ copies.  Since
we are dealing with a transformation from a tensor product of $N$
channels to a single channel,  the goal is to find the admissible
deterministic $(1^{\otimes N},1)$-map $\mathcal{R}$ which most faithfully
realizes the transformation $\bigotimes_{i=1}^{N} \mathcal{U}_g^{(i)}
\to \mathcal{V}_g $.
This can be expressed in terms of Choi operators as
\begin{align}
  R * \bigotimes_{i=1}^{N} \KetBra{U_g^{(i)}}{U_g^{(i)}} \simeq
\KetBra{V_g}{V_g}
\end{align}
where
$R \in \Lin( \bigotimes_{k=0}^{2N+1} \hilb{H}_{k} ) $ is a deterministic $(1^{\otimes N},1)$-comb
and we used Eq. \eqref{eq:choiunit} and Eq. \eqref{eq:linkchoi}.
It is worth stressing that, as we mentioned in
Section \ref{sec:highord}, such $\mathcal{R}$ does not necessarily
have a realization as a quantum circuit.
We now need a criterion to quantify how close
the channel
$ R * \bigotimes_{i=1}^{N} \KetBra{U_g^{(i)}}{U_g^{(i)}}$
is to the target $\KetBra{V_g}{V_g}$.
The closeness between two channels $\mathcal{C}, \mathcal{D} \in \Lin(\Lin(\hilb{H}_0),\Lin(\hilb{H}_1))$ can be expressed
in terms of the  channel fidelity \cite{raginski}, that is defined as
follows
\begin{align}
  \label{eq:chanfid}
  f(C,D) := \frac{1}{d_0^2} \left( \Tr\left[\sqrt{\sqrt{C}D\sqrt{C}}
    \right] \right) ^2 ,
\end{align}
where $C$ and $D$ are the Choi operators of the channels.
As a figure of merit for our task we use the channel fidelity between
$ R * \bigotimes_{i=1}^{N} \KetBra{U_g^{(i)}}{U_g^{(i)}}$
and $\KetBra{V_g}{V_g}$
 uniformly averaged over the unknown unitaries \cite{compact}, that is
 \begin{align}
&        F(R) = \int \! \!\d \! g \; f \! \left(R * \bigotimes_{i=1}^{N} \KetBra{U_g^{(i)}}{U_g^{(i)}} ,
\KetBra{V_g}{V_g}\right) =\nonumber \\
&=\frac{1}{d^2_{0}} \int \! \!\d \! g \Tr
\left[
  R \left(\KetBra{U_g^{*}}{U_g^{*}}\otimes
\KetBra{V_g}{V_g}
\right)
\right] \label{eq:fidelity1} \\
& U_g := \bigotimes_{i=1}^{N} U_g^{(i)} \; .
\nonumber
\end{align}

The problem we address consists of finding
a deterministic $(1^{\otimes N},1)$-comb $R$
that maximizes the function $F$ in Eq.
\eqref{eq:fidelity1}  i.e.
\begin{equation}
\begin{aligned}\label{eq:optprob1}
& \underset{R}{\text{maximize}}
& & F(R)\\
& \text{subject to}
& & R \mbox{ is a deterministic }(1^{\otimes N},1)\mbox{-comb.}
\end{aligned}
\end{equation}
Equation \eqref{eq:optprob1} can be formulated as a 
{\em semidefinite program},
 namely
a problem that can be phrased as
\begin{equation}
\begin{aligned}\label{eq:semidef}
& \underset{\rho}{\text{maximize}}
& & \Tr[\rho X]\\
&  \text{subject to}
& & \map F(\rho)\leq Y\\
& & & \rho\geq 0,
\end{aligned}
\end{equation}
where $X\in \Lin_{\defset{H}} (\hilb{H})$, $Y \in  \Lin_{\defset{H}}(\hilb{K})$, $\map
F:\Lin(\hilb{H})\to\Lin(\hilb{K})$, and
$\Lin_{\defset{H}}(\hilb{H})\subseteq\Lin(\hilb{H})$ denotes the space of
Hermitian operators on $\hilb H$, and the map $\map F$ is required to
be Hermitian-preserving.
The fact that the constraint ``$R$ is a deterministic $(1^{\otimes
  N},1)$-comb'' in Eq.~\eqref{eq:optprob1} involves equalities while
the constraint in Eq. \eqref{eq:semidef}
is given by the inequality $\map F(\rho)\leq Y$ does not
represent a problem. Indeed, one can easily see that for any
probabilistic $(1^{\otimes
  N},1)$-comb $R$ there exists a deterministic one $\overline{R}$
such that $F(\overline{R}) \geq F(R) $. For this reason,  we can
replace the optimization problem 
of Eq.  \eqref{eq:optprob1} with the following one 
\begin{align*}
  \begin{aligned}
& \underset{R}{\text{maximize}}
& & F(R)\\
& \text{subject to}
& & R \mbox{ is a probabilistic }(1^{\otimes N},1)\mbox{-comb}
\end{aligned}
\end{align*}
that is equivalent to a semidefinite programming
in the form of Eq. \eqref{eq:semidef}.

In the next subsections we will see that by exploiting symmetries it
is possible to radically simplify the problem, reducing it to a much
simpler semidefinite program.

\subsection{Optimality of the parallel strategy} \label{parallelrules}

As we discussed at the end of Section \ref{sec:highord} the set of
admissible $(1^{\otimes N},1)$-maps includes mathematical objects that
currently lack a physical interpretation.  Before dealing with the
optimization problem it is good to know whether the map which
maximizes Eq.~\eqref{eq:fidelity1} is known to be realizable in the
physical world.  In this subsection we prove that the symmetries of
the problem allow us to choose the optimal map $\mathcal{R}$ to be a
deterministic $2$-comb. This fact by Proposition
\ref{prop:realization} implies that $\mathcal{R}$ can be realized as a
concatenation of channels and the task can be optimally accomplished
using quantum circuits.  We start by
proving the following lemma.

\begin{lemma} \label{lem:covacomb}
  The optimal deterministic $(1^{\otimes N},1)$-comb $R \in \Lin (\bigotimes_{k=0}^{2N+1} \hilb{H}_{k})$
which maximizes Eq.~\eqref{eq:fidelity1}
can be assumed without loss of generality
to satisfy the commutation relation
\begin{align} \label{eq:covacomb}
  \begin{split}
  & [R, W^*_h \otimes W_g]=0 \qquad \forall g,h \in G,
  \end{split}
  \end{align}
where
$W_{h} = U^*_{h} \otimes V_h $ with
$U_{h} \in \bigotimes_{i=1}^N\Lin(\hilb{H}_{2i-1})$,
and $V_h \in \Lin(\hilb{H}_{0})$, and
$W_{g} = U^*_{g} \otimes V_g $ with
 $ U_{g} \in \bigotimes_{i=1}^N \Lin(\hilb{H}_{2i})$,
$V_g\in \Lin(\hilb{H}_{2N+1})$.
\end{lemma}

\Proof
The proof follows the Holevo's averaging argument for covariant
estimation \cite{holevo}.
 Let $R$ be optimal.
Then consider the operator
\begin{align*}
  &\tilde{R} :=
\int \d \! g \; \d \!h \,( W^*_h\otimes W_g )R( W^{T}_h \otimes W_g^\dag).
\end{align*}

The set of deterministic
$(1^{\otimes N},1)$-combs is a convex set,
hence $\tilde{R}$ is a well defined
 deterministic
$(1^{\otimes N},1)$-comb.
One can easily verify that
$\tilde{R}$ satisfies Eq. \eqref{eq:covacomb}
and $F(R) = F(\tilde{R})$.
\qed

Lemma  \ref{lem:covacomb} is the key ingredient
for proving the following proposition.

\begin{proposition}\label{prop:parallelism}
Let $R$ be a  $(1^{\otimes N},1)$-comb in $\Lin (\bigotimes_{k=0}^{2N+1}
 \hilb{H}_{k})$ which obeys the commutation relation
\eqref{eq:covacomb}.
Then there exist a deterministic $2$-comb $R'$ formed by
channels
$\mathcal{C}_1 : \Lin(\hilb{H}_0) \to \Lin(\bigotimes_{i=1}^N
\hilb{H}_{2i-1} \otimes \hilb{H}_M)$,
and
$\mathcal{C}_2 : \Lin(\bigotimes_{i=1}^N
\hilb{H}_{2i} \otimes \hilb{H}_M) \to \Lin(\hilb{H}_{2N+1}) $,
such that
\begin{align}
  \label{eq:1}
R *\KetBra{U_g}{U_g} &=R' *\KetBra{U_g}{U_g} \nonumber \\
  &=     C_1 *
\KetBra{U_g}{U_g}
* C_2
\quad \forall g\in G,
\end{align}
\end{proposition}
where $R'=C_1*C_2$ and the link is performed on $\hilb{H}_M$.

\Proof Let $R$ be a $(1^{\otimes N},1)$-comb in $\Lin
(\bigotimes_{k=0}^{2N+1} \hilb{H}_{k})$ and let us define $\hilb{H}_A
= \bigotimes_{i=1}^N \hilb{H}_{2i-1}$ and $\hilb{H}_B =
\bigotimes_{i=1}^N \hilb{H}_{2i}$.  With this notation we have $R \in
\Lin ( \hilb{H}_0 \otimes\hilb{H}_A \otimes \hilb{H}_B \otimes
\hilb{H}_{2N+1})$ and $U_g \in \Lin ( \hilb{H}_B)$.  Let us consider
the operator $S \in \Lin ( \hilb{H}_0 \otimes\hilb{H}_A \otimes
\hilb{H}_B)$ defined as $S := \Tr_{2N+1}[R]$.  Upon introducing
auxiliary hilbert spaces $\hilb{H}_{0'} \equiv \hilb{H}_{0} $, $
\hilb{H}_{A'} \equiv \hilb{H}_{A}$ and $ \hilb{H}_{B'} \equiv
\hilb{H}_{B}$, it is possible to define the rank one operator
$\KetBra{S^{\frac12}}{S^{\frac12}} \in \Lin ( \hilb{H}_0
\otimes\hilb{H}_A \otimes \hilb{H}_B \otimes \hilb{H}_E)$, where
$\KetBra{S^{\frac12}}{S^{\frac12}} = (S^{\frac12} \otimes I_E)
\KetBra{I}{I} (S^{\frac12} \otimes I_E)$ and we also defined
$\hilb{H}_E := \hilb{H}_{0'} \otimes \hilb{H}_{A'} \otimes
\hilb{H}_{B'}$, and identity $I_E$ on $\hilb{H}_E $.  The commutation
\eqref{eq:covacomb} implies $[S^{\frac12}, I_0 \otimes I_{A} \otimes
U_g]=0$, and together with Eqs.~\eqref{eq:doubleket},
\eqref{eq:choiunit} and \eqref{eq:link} we have
\begin{align}
  \begin{split}
  \KetBra{S^{\frac12}}{S^{\frac12}} *
  \KetBra{U_g}{U_g}= \\
  U_g
\left( \KetBra{S^{\frac12}}{S^{\frac12}} *  \KetBra{I}{I} \right)
U^\dag_g \, , \label{eq:chanc1}
  \end{split}
\end{align}
where $\KetBra{U_g}{U_g}\in\hilb{H}_B \otimes \hilb{H}_A$, and in the
last line $U_g \in \Lin( \hilb{H}_{B'})$.  From the definition of a
$(1^{\otimes N},1)$-comb we have that $R * \KetBra{I}{I}$ is a channel
from $ \Lin( \hilb{H}_{0}) $ to $ \Lin( \hilb{H}_{2N + 1}) $ and then
$\Tr_{2N+1}[R * \KetBra{I}{I}] = I_0$.  From this relation, from
$\Tr_E \left [\KetBra{S^{\frac12}}{S^{\frac12}} \right] = S$ and from
the definition of $S$ we have
\begin{align}
&\Tr_E \left[ \KetBra{S^{\frac12}}{S^{\frac12}} *  \KetBra{I}{I} \right] = \nonumber\\
&= S *  \KetBra{I}{I}  =
\Tr_{2N+1}[ R *  \KetBra{I}{I} ] = I_0.
\label{eq:prev}
\end{align}
Denoting by $C_1$ the CP map with Choi operator
$C_1:=\KetBra{S^{\frac12}}{S^{\frac12}} * \KetBra{I}{I}$, one can
easily realize that by virtue of Eq.~\eqref{eq:prev} $C_1$ is a channel
from $\Lin(\hilb{H}_0)$ to $\Lin (\hilb{H}_{E})$.
Eq.~\eqref{eq:chanc1} can be diagrammatically represented as
\begin{align}\label{eq:halfproofpar}
  U_g
C_1
U^\dag_g
=
C_1 * \KetBra{  U_g}{  U_g} =
   \begin{aligned}
    \Qcircuit @C=0.8em @R=1.1em {
    \ustick{0}  &\multigate{1}{\mathcal{C}_1}&\ustick{B'}\qw &\gate{\mathcal{U}_g}&
\ustick{B'}\qw \\
&\pureghost{C_1} &  \qw &\ustick{M} \qw &\qw
}
  \end{aligned},
\end{align}
where we defined $\hilb{H}_M := \hilb{H}_{0'} \otimes \hilb{H}_{A'}$.

Let us now introduce the operator $C_2 := T S^{-\frac12} R
S^{-\frac12} T^\dag+(I-\Pi_S)_E\otimes \tfrac1{d_{2N+1}}I_{2N+1}$,
where $T$ is the isomorphism between $\hilb{H}_0 \otimes \hilb{H}_A
\otimes \hilb{H}_B $ and $\hilb{H}_{0'} \otimes \hilb{H}_{A'} \otimes
\hilb{H}_{B'} $, and $\Pi_S:=S^{-\frac12}SS^{-\frac12}$ is the
projection on the support of $S$. Since we have that $\Tr_{2N+1} [ T
S^{-\frac12} R S^{-\frac12} T^\dag] = \Pi_S$, it is easily verified
that $C_2$ is a channel from $\hilb{H}_E$ to $\hilb{H}_{2N+1}$.  By
direct computation one can verify that
\begin{align}\label{eq:linkcis}
  \KetBra{S^{\frac12}}{S^{\frac12}} * C_2 = R,
\end{align}
where the link is performed on the Hilbert space $\hilb{H}_{E}$.
Combining Eqs. \eqref{eq:chanc1}, \eqref{eq:halfproofpar} and \eqref{eq:linkcis}
and exploiting commutativity and associativity of the link product we get
\begin{align}
  \begin{split}
       R *  \KetBra{  U_g}{  U_g} &=
  \KetBra{S^{\frac12}}{S^{\frac12}} * C_2 *     \KetBra{
    U_g}{  U_g} = \\
&= C_1 *   \KetBra{U_g}{  U_g} * C_2 . \nonumber
  \end{split}
\end{align}
which is Eq. \eqref{eq:1}
up to relabeling of Hilbert spaces.
\qed

Proposition \ref{prop:parallelism} tells us that the optimal transformation
from a set of unitary transformations
 $\{ \mathcal{U}^{(i)}_g \}_{i=1 \dots N}$
to a target unitary channels
$\mathcal{V}_g$ is physically realizable with the following scheme:

\begin{enumerate}[i.]
\setlength{\itemsep}{1pt}
\setlength{\parskip}{0pt}
\setlength{\parsep}{0pt}
\item application of a preprocessing channel $\mathcal{C}_1$
from $\hilb{H}_0$ to $(\bigotimes_{i} \hilb{H}_{2i-1}) \otimes \hilb{H}_M$;
\item parallel application of the
unitary channels $\mathcal{U}^{(i)}_g$ on $\hilb{H}_{2i-1}$;
\item final application of a postprocessing channel
 $\mathcal{C}_2$ from
$(\bigotimes_{i} \hilb{H}_{2i-1}) \otimes \hilb{H}_M$ to
$\hilb{H}_{2N + 1}$.
\end{enumerate}
This means that the problem of finding an optimal $(1^{\otimes N},1)$-comb mapping the set
of unitary channels
 $\{ \mathcal{U}^{(i)}_g \}_{i=1 \dots N}$
to
$\mathcal{V}_g$
is equivalent to the problem of finding an
optimal $2$-comb that maps
a single channel
$\mathcal{U}_g = \bigotimes_{i=1}^N
\mathcal{U}^{(i)}_g$
to $\mathcal{V}_g$,
\begin{align}
\begin{aligned}
    \Qcircuit @C=0.7em @R=0.4em {
      &\ustick{0}&\multigate{4}{\mathcal{C}_1}&\ustick{1} \qw&\gate{\mathcal{U}^{(1)}_g}& \ustick{2}\qw & \multigate{4}{\mathcal{C}_2} \qw&\ustick{2N+1} \qw\\
      & &\pureghost{\mathcal{C}_1}&\ustick{3}\qw&\gate{\mathcal{U}^{(2)}_g}&\ustick{4}\qw&  \ghost{\mathcal{C}_2}& \\
      & &\pureghost{\mathcal{C}_1}&  &     \vdots   &   & \pureghost{\mathcal{C}_2}& \\
      & &\pureghost{\mathcal{C}_1}&  &                  &   & \pureghost{\mathcal{C}_2}& \\
       & &\pureghost{\mathcal{C}_1}&\qw &\ustick{M}\qw          &\qw  & \ghost{\mathcal{C}_2}& }
  \end{aligned}
\to
\begin{aligned}
     \Qcircuit @C=0.6em @R=1.1em {
    \ustick{0}  &\multigate{1}{\mathcal{C}_1}&\ustick{1}\qw &\gate{\mathcal{U}_g}&
\ustick{2}\qw& \multigate{1}{\mathcal{C}_2}& \ustick{3} \qw \\
&\pureghost{\mathcal{C}_1} &  \qw &\ustick{} \qw &\qw  & \ghost{\mathcal{C}_2}&
} \nonumber
\end{aligned},
\end{align}
where we made a suitable relabeling of the Hilbert spaces.

\subsection{The optimal circuit}
\label{sec:optcircuit}
Thanks to the results of the previous section,
the optimization problem \eqref{eq:optprob1}
can be restated as follows:

\begin{align}
&\underset{R}{\text{maximize}} && F(R)= \frac{1}{d_0^2} \!
\int \d \! g
\Bra{V_g}_{30} \Bra{U_g^*}_{21} R \Ket{U_g^*}_{21}
\Ket{V_g}_{30}\nonumber \\
&\text{subject to} && \Tr_3[R] = I_2 \otimes S_{10},\ \Tr_1[S] = I_0,\ R, S \geq 0  \label{eq:optprob2}
\end{align}
where we used the notation $\Ket{A}_{ij} \in \hilb{H}_{i} \otimes
\hilb{H}_j$.  The constraints on $R$ translate the condition that $R$
is a deterministic $2$-comb (see Eq. \eqref{Eq:normcomb}).

As a consequence of Lemma \ref{lem:covacomb} it is not restrictive to
search for the optimal comb $R$ for the problem (\ref{eq:optprob2})
among those having the following symmetry
\begin{align} \label{eq:covacomb1}
  \begin{split}
    & [R , (V_h^* \otimes U_h)_{01} \otimes (U_g^* \otimes V_g)_{23} ]
    =0 \qquad \forall g,h \in G,
  \end{split}
\end{align}
where the two independent unitary representations of group $G$ that
act on Hilbert spaces $\hilb{H}_0 \otimes \hilb{H}_1$ and $\hilb{H}_2
\otimes \hilb{H}_3$, respectively. 
It is now useful to consider the decompositions of $U$ and $V$ into
irreducible representations as follows
\begin{align}
U_h= \bigoplus_{\beta} U^{[\beta]}_h  \otimes I_{m_\beta}
\qquad
V_h= \bigoplus_{a} V^{[a]}_h \otimes I_{m_a} \nonumber\\
U_g= \bigoplus_{\gamma} U^{[\gamma]}_g  \otimes I_{m_\gamma}
\qquad
V_g= \bigoplus_{d} V^{[d]}_g \otimes I_{m_d},\nonumber
\end{align}
where for
$\forall f\in G$, $x\in\{a,d\}$, $\xi\in \{\beta, \gamma\}$
\begin{align}
U^{[\xi]}_f \in \Lin(\hilb{H}_\xi) \quad V^{[x]}_f \in \Lin(\hilb{H}_x)
\end{align}
are unitary irreducible representations (irreps) of $G$ and $I_\xi$,
$I_{m_x}$ are the identity operators on the multiplicity spaces
$\hilb{H}_{m_\xi} $ and $\hilb{H}_{m_x}$.  As we prove in appendix
\ref{a:multiplicity}, we can without loss of generality restrict
ourselves to the case in which the multiplicity spaces
$\hilb{H}_{m_\xi}$ 
are one dimensional for all $\xi$ i.e.  $U_f= \bigoplus_{\xi}
U^{[\xi]}_f$ and
\begin{align}
\label{decomporep}
 V^{[x]}_f \otimes U^{[\xi]*}_f = \bigoplus_{Y}
 W^{[Y]}_f \otimes I_{m^{x,\xi}_Y} \; .
\end{align}
Eq. \eqref{decomporep} induces the following decomposition of
Hilbert spaces:
\begin{align} \label{eq:decompohilbspace}
  \begin{aligned}
    \hilb{H}_0 = \bigoplus_a \hilb{H}_a \otimes \hilb{H}_{m_a} \quad
    \hilb{H}_1= \bigoplus_\beta \hilb{H}_\beta
\\
         \hilb{H}_2 = \bigoplus_\gamma \hilb{H}_\gamma  \quad
         \hilb{H}_3= \bigoplus_d \hilb{H}_d \otimes
         \hilb{H}_{m_d}
\\
         \hilb{H}_0 \otimes  \hilb{H}_1 =
         \bigoplus_{K} \hilb{H}_K \otimes
         \hilb{H}_{m_K}
\\
         \hilb{H}_2 \otimes  \hilb{H}_3 =
         \bigoplus_{L} \hilb{H}_L \otimes
         \hilb{H}_{m_L}
\\
\hilb{H}_{m_K} = \bigoplus_{a, \beta} \hilb{H}_{m^{a, \beta}_K}
\quad
\hilb{H}_{m_L} = \bigoplus_{\gamma ,d} \hilb{H}_{m^{\gamma,d}_L}
  \end{aligned}
\end{align}

The commutation relation \eqref{eq:covacomb1} can be rewritten as
\begin{align}\label{eq:covacomb2}
\left[R,  \left( \bigoplus_K  W^{[K]}_h \otimes I_{m_K}\right) \otimes \left(\bigoplus_L
  W^{*[L]}_g \otimes I_{m_L} \right) \right] =0 
\end{align}
which thanks to the Schur lemma's implies
\begin{align}
  \label{eq:3}
  R = \sum_{K,L} \Pi^K \otimes \Pi^L \otimes R^{KL}
\end{align}
where $R^{KL}\in \mathcal{L}(\hilb{H}_{m_K} \otimes \hilb{H}_{m_L})$ and $\Pi^Y$ for $Y\in\{K,L\}$ is a projector onto $\hilb{H}_Y$.
It is convenient to define the projectors
 $P^{x,\xi}_Y$ on the multiplicity space $\hilb{H}_{m^{x,\xi}_Y}$,
$P_x$ on the multiplicity space
$\hilb{H}_{m_{x}}$ and
the $\Pi_x$ on the representation spaces
$\hilb{H}_{x}$.
We also define projector $P^x_Y := \sum_\xi P^{x,\xi}_Y$ onto a subspace  $\bigoplus_{\xi} \hilb{H}_{m^{x, \xi}_Y}$.
In the following
 $m_x$ will denote the dimension of the
multiplicity space $\hilb{H}_{m_x}$,
$d_x$ will denote the dimension of the representation
space $\hilb{H}_{x}$ and
$m^{x,\xi}_Y$ will denote the dimension of
the multiplicity space $\hilb{H}_{m^{x,\xi}_Y}$.

The main result of this section, stated in the following proposition,
is that the optimization problem~\eqref{eq:optprob2} can be
transformed into an optimization problem defined by a set of quadratic
expressions for a probability distribution vector.

\begin{proposition}\label{prop:optimality}
  Let us consider the following optimization problem
\begin{align}
  \begin{aligned} \label{eq:optprogfinal}
    & \underset{p_K^a}{\mbox{\rm maximize}}
& & \Phi(p_K^a)=
\sum_K \left( \sum_a \sqrt{q_K^a p_K^a} \right)^2
 \\
& \mbox{\rm subject to}
& & \sum_K p_K^a =1 \quad \forall a \\
&&& p_K^a \geq 0 \;.
  \end{aligned}
\end{align}
where $q^a_K = \frac{m_a d_a}{d_K d^2_{0}}\sum_\beta  m^{a\beta}_K
d_\beta$ and let  $\check{R} = \check{R}(p_K^a)$
be defined as follows:
\begin{align}
  \begin{aligned}
    \label{eq:optimalcomb}
&\check{R}  := \sum_{KL} \Pi_K \otimes \Pi_L \otimes
\check{R}^{KL} \\
&\check{R}^{KL} =
\left(\delta_{KL} \ket{\psi_K}\bra{\psi_K}+\sum_{\beta} D^\beta_K \otimes \sum_{\gamma\neq \beta} \Delta^{\gamma}_{L}\right) \\
&D^\beta_K=d_K d_\beta\sum_a p^{a}_{K} \frac{ P^{a \beta}_K} {h^a_K}
\qquad \Delta^{\gamma}_{L}=\frac{d_{\gamma} P^{\gamma}_L} {\Tr[P^{\gamma}_L] d_L k^{\gamma}_L}\\
&\ket{\psi_K}=\sum_{a, \beta}\sqrt{\frac{p^a_K d^2_\beta} {h^a_K}}
\Ket{I_{m^{a,\beta}_K}} \\
&h^a_K=\frac{d^2_K} {m_a d_a}\sum_\beta  m^{a\beta}_K d_\beta ,
  \end{aligned}
\end{align}
where $k^{\gamma}_L$ denotes for how many $L$'s $W^{[L]}_g$ is in the
decomposition of $U_g^{*[\gamma]} \otimes V_g^{[d]}$ for some $d$.

If $\tilde{p}_K^a$
is a solution of the optimization problem \eqref{eq:optprogfinal}
then $\check{R}(\tilde{p}_K^a)$ is a solution of the optimization problem
of \eqref{eq:optprob2} and $F(\check{R}) = \Phi(\tilde{p}_K^a)$.
\end{proposition}

We split the proof of Proposition \ref{prop:optimality}
into two parts.
In the first lemma we prove that
the operator defined through the ansatz
of Eq. \eqref{eq:optimalcomb} is
a well defined deterministic $2$-comb.

\begin{lemma}
  \label{th:normalization1}
Let
 $\check{R}$ have the form as in Eq. \eqref{eq:optimalcomb}.
 Then $\check{R}$  satisfies the constraints
of Eq. \eqref{eq:optprob2} if and only if $\sum_{K} p^{a}_{K}=1 \; \forall a$
and $p^{a}_{K} \geq 0$.
\end{lemma}

\Proof
We will utilize identities
\begin{align}
  \begin{aligned}
    \label{eq:defproj}
&(\Pi_a \otimes P_{a})_{0}\otimes I_1 = \sum_{K,\beta} \Pi^K \otimes P^{a,\beta}_K \\
&(\Pi_\gamma)_2 \otimes I_3 = \sum_{L,d} \Pi^L \otimes P^{\gamma,d}_L
  \end{aligned}
\end{align}
that follow from the decompositions \eqref{eq:decompohilbspace}.

We recall the normalization constraints for a $2$-comb $R$:
$\Tr_{3}[{R}]=I_{2}\otimes S_{10}$ and $\Tr_{1}[{S}]=I_0$.
If $R$ obeys Eq. (\ref{eq:covacomb1}) we have
$\Tr_{3}[R] = \sum_\gamma \Pi_\gamma \otimes \sum_K \Pi^K \otimes
S^{K,\gamma}$ and the condition $\Tr_{3}[{R}]=I_{2}\otimes S_{10}$
is equivalent to 
\begin{align}
 \label{eq:normcond1i}
 & \sum_{L,d} \frac{d_L}{d_\gamma} \Tr_{m_L}\left({R}^{KL}\; P^{\gamma,d}_L\right)={S}^{K} \quad \forall \gamma ,\forall K,
\end{align}
where $\Tr_{m_L}$ indicates the trace over $\hilb{H}_{m_L}$.
Similarly, the condition $\Tr_{1}[{S}]=I_0$ can be rewritten as
\begin{align}
\label{eq:normcond2}
d_a P_a = \sum_{K,\beta} \Tr_{\hilb{H}_a}\Tr_{1}\left[{S} (\Pi^K \otimes
  P^{a,\beta}_K) \right] \quad \forall a.
\end{align}
We notice that by construction operator $\check{R}$ defined in Eq.~\eqref{eq:optimalcomb} obeys the symmetry from Eq. (\ref{eq:covacomb1}).
In addition $\check{R}$ obeys $\Tr_{23}[\check{R}]/ \dim \hilb{H}_2=\check{S}_{10}=\sum_{K,a,\beta} s^{a,\beta}_K \Pi^K \otimes  P^{a,\beta}_K$, which allows us to rewrite Eqs. (\ref{eq:normcond1i}) and \eqref{eq:normcond2} for $\check{R}$ in more convenient form:
\begin{align}
  \label{eq:normcond1f}
 & \sum_{L,d} \frac{d_L}{d_\gamma} \Tr_{m_L}\left(\check{R}^{KL}\; P^{\gamma,d}_L\right)=\check{S}^{K} \quad \forall \gamma ,\forall K, \\
\label{eq:normcond2f}
&\sum_{K,\beta} \frac{d_K}{d_a m_a} \Tr_{ m_K}\left(\check{S}^K \; P^{a,\beta}_K
\right)=1.  \quad \forall a,
\end{align}
where we used Eq. (\ref{eq:defproj}).
In the following we demonstrate that the above two equations 
are fulfilled, i.e. $\check{R}$ is a properly normalized $2$-comb.
We notice that that
$\Tr_{m_L}[\ket{\psi_K}\bra{\psi_K} P^{\gamma}_K] = d_\gamma
D^{\gamma}_K /d_K$, $\Tr_{m_L}[\Delta^{\gamma'}_L P^{\gamma}_L] =
\delta_{\gamma \gamma'} d_\gamma /(k^\gamma_L d_L)$.  This implies
$(d_L /d_\gamma) \sum_{L} \Tr_{m_L}[\check{R}^{KL}\; P^{\gamma}_L]=
\sum_{\beta} D^{\beta}_K \equiv \check{S}^K$ for each $K$ and
independently on $\gamma$. Thus, the first normalization condition is
satisfied.  Inserting $\check{S}^K$ into Eq. (\ref{eq:normcond2f})
we obtain the condition $\sum_{K} p^{a}_{K}=1 \; \forall a$.  The
positivity of the $p_K^a$ guarantees the positivity of
$\check{R}$ and $\check{S}$.
\qed

In the next lemma we prove that the
deterministic $2$-comb that solves the optimization problem
\eqref{eq:optprob2} can be assumed without loss of generality
to be of the form of Eq. \eqref{eq:optimalcomb}.

\begin{lemma}
  For any deterministic $2$-comb $R$ there exist
a set of positive coefficients $p_K^a$, $\sum_K^a p_K^a =1$ $\forall
a$ such that  for
the  $2$-comb $\check{R}(p_K^a)$ defined by
Eq.~\eqref{eq:optimalcomb}
we have $F(R) \leq F(\check{R}) = \Phi(p_K^a)$.
\end{lemma}
\Proof
From Eqs.  \eqref{eq:covacomb2} and \eqref{eq:3}  we have
\begin{align}
  \begin{aligned}
    \label{eq:fmstep2}
F(R)= \frac{1}{d_0^2} \!
\int \d \! g
\Bra{V_g}_{30} \Bra{U_g^*}_{21} R \Ket{U_g^*}_{21}
\Ket{V_g}_{30} =\\
=\frac{1}{{(d_{0})}^2} \Bra{I}_{30} \Bra{I}_{21} R \Ket{I}_{21}
\Ket{I}_{30}= \\
 =\sum_K \frac{d_K}{{(d_{0})}^2} \sum_{a,\beta,\gamma,d} \bb I_{m^{a,\beta}_K}|R^{KK}|I_{m^{\gamma,d}_K}\kk,
  \end{aligned}
\end{align}
where $|I_{m^{a,\beta}_K}\kk \in \hilb{H}_{m^{a,\beta}_K}\otimes \hilb{H}_{m^{a,\beta}_K}$ and we used
$|I\kk_{03}|I\kk_{12}=\sum_K |I_{K} \kk \sum_{a,\beta} |I_{m^{a,\beta}_K}\kk$
with $|I_{K} \kk \in \hilb{H}_K \otimes \hilb{H}_K$.

For a positive operator $X$ and arbitrary vectors $\ket{\psi}$ and
$\ket{\phi}$
we have:
$|\bra{\psi}X\ket{\varphi}|\leq\sqrt{\bra{\psi}X\ket{\psi}}\sqrt{\bra{\varphi}X\ket{\varphi}}$,
$\bra{\psi}X\ket{\psi}\leq \bra{\psi} \psi \rangle  \; \Tr[X]$.
Moreover we have $ \BraKet{I_{m^{a,\beta}_K}}{I_{m^{a,\beta}_K}}= m^{a,\beta}_K$ and
$\Ket{I_{m^{a,\beta}_K}} = P^{a,\beta}_K \otimes P^{a,\beta}_K \Ket{I_{m^{a,\beta}_K}}$.
Applying the above two inequalities
to Eq. (\ref{eq:fmstep2}) we obtain
\begin{align}
\label{eq:ineqfom1}
\begin{aligned}
&  F(R)\leq \sum_K \frac{d_K}{{(d_{in})}^2}\left( \sum_{a,\beta} \sqrt{m^{a,\beta}_K d_\beta} \sqrt{\frac{R^{a\beta \beta a}_{KK}}{d_\beta}}\right)^2 \leq\\
&\leq\sum_K \frac{d_K}{{(d_{in})}^2}\left( \sum_{a}\sqrt{\sum_{\beta}
    m^{a,\beta}_K d_\beta \sum_{\beta'} \frac{R^{a\beta' \beta'
        a}_{KK}}{d_{\beta'}}}\right)^2\leq
\\
&\leq \sum_K \frac{d_K}{{(d_{in})}^2}\left( \sum_{a}\sqrt{\sum_{\beta}
    m^{a,\beta}_K d_\beta \sum_{\beta' L d} \frac{d_L R^{a\beta' \beta'
        d}_{KL}}{d_{\beta'} d_K}}\right)^2
\end{aligned}
\end{align}
where we used Schwarz inequality again in the second step
and we defined $R_{KL}^{a \beta \gamma d} = \Tr[R^{KL} P_K^{a,\beta}
\otimes P_L ^{\gamma, d}]$.
Let us now define
$p^a_K = \sum_{\beta, L, d} (d_K d_L R_{KL}^{a \beta \beta d})/(m_a
d_a d_\beta)$ and $\check{R}(p_K^a)$ with the ansatz of
Eq. \eqref{eq:optimalcomb}.
We notice that the positivity of
$R_{KL}^{a \beta \beta d}$ implies that $p_K^a \geq 0$.
By substituting the above definition
into Eq.~\eqref{eq:ineqfom1}
we have
\begin{align*}
 F(R) \leq \sum_K\left( \sum_a \sqrt{q_K^a p_K^a} \right)^2 = F(\check{R})
\end{align*}
where we inserted the definition of $q_K^a$
given in Proposition~\ref{prop:optimality}.

It only remains to prove that  $\sum_K p_K^a=1$ $\forall a$.
Since Eq. (\ref{eq:normcond1f}) holds for any $\gamma$ we can insert
Eq. (\ref{eq:normcond1f}) into Eq. (\ref{eq:normcond2f}) in such a way that for every term we choose
$\gamma=\beta$ and obtain
\begin{align*}
  \sum_{K,\beta,L,d} \frac{d_L d_K}{d_a m_a d_\beta}\Tr[R^{KL} P_K^{a,\beta}
\otimes P_L ^{\beta, d}] =1 \quad \forall a
\end{align*}
which completes the proof.\qed

One can now easily prove that the problem in
Eq.~\eqref{eq:optprogfinal} can be expressed as a semidefinite program
of Eq.~\eqref{eq:semidef}. Indeed, one can take the spaces $\hilb
H:=\operatorname{span}(\ket{a}\otimes\ket{K})$ and $\hilb
K:=\operatorname{span}(\ket{a})$, with
$X:=\sum_{K}\ketbra{\varphi_K}{\varphi_K}\otimes \ketbra{K}{K}$,
$\ket{\varphi_K}:=\sum_{a}\sqrt{q^a_K}\ket{a}$,
$Y:=\sum_a\ketbra{a}{a}$, the map $\map F$ being just given by
\begin{equation}
  \map F(\rho):=\sum_{K,a} \Tr[\rho(\ketbra{a}{a}\otimes\ketbra{K}{K})]\ketbra{a}{a}.
\end{equation}
Finally, notice that the constraint in Eq.~\eqref{eq:optprogfinal}
involves an equal sign, namely $\map F(\rho)=Y$. However, we can
without loss of generality consider the looser constraint
$\map F(\rho) \leq Y$ because for any $\rho$ satisfying $\map F(\rho)<
Y$ one can find $\rho'$ such that $\map F(\rho')=Y$ and
$\Tr[\rho'X] \geq \Tr[\rho X]$. This implies that the final
formulation corresponds to a much simpler semidefinite program than
the original one in Eq.~\eqref{eq:optprob1}.

\section{Examples}
\label{sec:examples}
\subsection{Transformations between irreducible representations}
\begin{figure}[h]
  \begin{center}
    \includegraphics[width=8cm ]{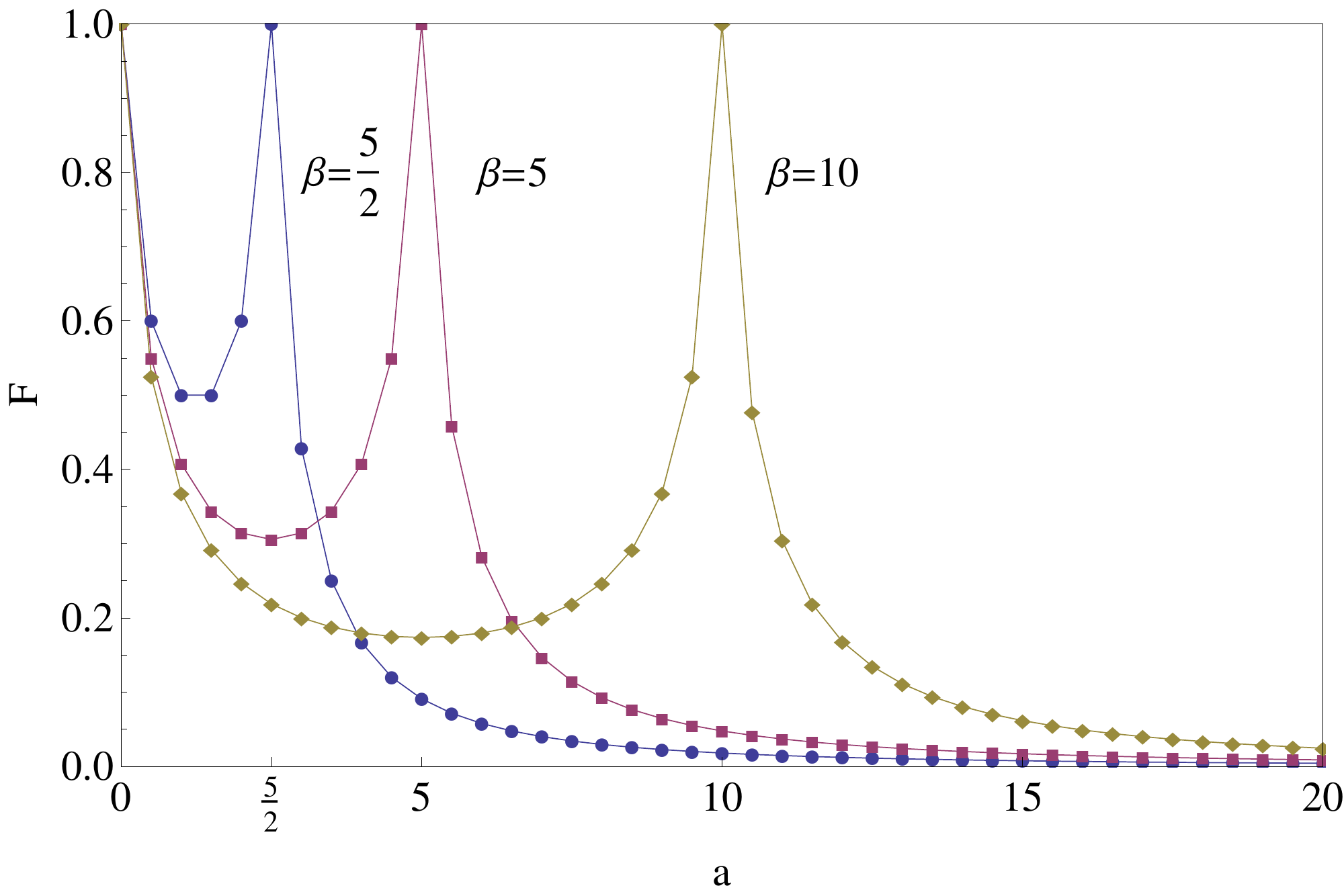}
    \caption{Average process fidelity $F$ of an optimal transformation
      between irreducible representations $\beta$ and $a$ for the
      group $SU(2)$. The three lines correspond to different choice of
      the starting representation $\beta$, while the $x$ axes
      represents the choice of the target irrep $a$.}
    \label{fig:transfgraph}
  \end{center}
\end{figure}

The simplest problem that falls into our general setting is the
transformation of unitary channels from an irreducible representation
$\beta$ of group $SU(2)$ into channels from a different irreducible
representation $a$ of the same group.  Since we have only one irrep
$a$ the figure of merit (\ref{eq:optprogfinal}) simplifies to
$F(p^a_K)=\sum_K q^{a}_K p^{a}_{K}$.  It is clear that the maximum
$F=\max_K q^a_K$ is achieved by a probability distribution $p^a_K$
with just one non zero entry.  Let us remind that the irreps of
$SU(2)$ are defined by a half-integer called \emph{spin}, and the
generators of the representation with spin $l$ are the usual quantum angular
momentum components $J^{(l)}_x,J^{(l)}_y,J^{(l)}_z$. Notice also that
for the group $SU(2)$ the complex conjugate representation of spin $l$
is equivalent to the $l$ representation, and is obtained by
conjugating the $l$ representation with the unitary $\exp(-i\pi
J_y)$. Moreover, the irreps of $SU(2)$ obey a simple composition rule,
when they are tensorized
\begin{align}
\label{eq:spindec}
U_a\otimes U_\beta = \bigoplus_{K=|a-\beta|}^{a+\beta} U_K
\end{align}
This implies $m^{a \beta}_K=1$ and in our case for each $K,a$ there
exists exactly one $\beta$, which leads with $a$ to irrep $K$. Since the
dimension of the spin $j$ irrep is $d_j=2j+1$ we have $q^a_K
=d_\beta/(d_K d_a)=(2\beta+1)(2a+1)^{-1}(2K+1)^{-1}$ and
\begin{align}
F_{max}=\frac{2\beta+1}{(2a+1)(2|a-\beta|+1)}.
\end{align}
As one might expect we can mimic reasonably only the irreps that have
spin number $a$ very close to $\beta$, the irrep that we have at
disposal, or irreps that are very close to the trivial representation,
namely those having a very small $a$. For illustration of the
achievable process fidelities see Fig. \ref{fig:transfgraph}.

\subsection{$1\rightarrow 2$ Cloning of $SU(d)$ transformations}

The results of section \ref{sec:optcircuit} enable us to simplify the
optimization of the $1\rightarrow 2$ cloning of a $SU(d)$
transformation originally done in Ref.~\cite{cloning}. From our
current perspective the problem might be viewed as a transformation of
the defining representation $U$ of $SU(d)$ into the reducible
representation $U\otimes U$.  The $d_0=d^2$ dimensional representation
$U\otimes U$ decomposes into two irreps acting in symmetric and
antisymmetric subspaces $\hilb{H}_\pm$. Let us distinguish these
irreps by the index $a=\pm$. Their dimensions are $d_{\pm}=d(d\pm
1)/2$. On the other hand, the starting representation $U$ is
irreducible, which implies that the indices $\beta$ and $\gamma$ range
over a single value. The representation $U\otimes U \otimes U^*$
defining the symmetries in Eq.  (\ref{eq:covacomb1}) decomposes into
three irreps, which we denote $\hat{\alpha}, \hat{\beta},
\hat{\gamma}$. The $d$-dimensional representation $\hat{\alpha}$
appears with multiplicity two, whereas $\hat{\beta}$ and
$\hat{\gamma}$ have multiplicity one and dimensions $d(d_+ -1)$,
$d(d_- -1)$, respectively.

The following table summarizes all the parameters $q^a_K$ that are
used for the optimization in Proposition \ref{prop:optimality}. For
the sake of simplicity we actually report the expressions for
$d^4q^a_K$.
\begin{center}
  \begin{tabular}{|r|c|c|c|}
    \hline
    & $\;K=\hat{\alpha}\;$ & $K=\hat{\beta}$ & $K=\hat{\gamma}$ \\
    \hline 
    $a=+$ & $\;\;d_+\;\;$ & $d_+ /(d_+ - 1)$ & $0$ \\
    \hline
    $a=-$ & $d_-$ & $0$ & $d_- /(d_- - 1)$ \\
    \hline
  \end{tabular}.
\end{center}
The figure of merit (\ref{eq:optprogfinal}) for this problem
then takes the following form
\begin{align}
  F=\left(\sqrt{q^+_{\hat{\alpha}}
      p^+_{\hat{\alpha}}}+\sqrt{q^-_{\hat{\alpha}}
      p^-_{\hat{\alpha}}}\right)^2 +
  q^+_{\hat{\beta}}(1-p^+_{\hat{\alpha}}) +
  q^-_{\hat{\gamma}}(1-p^-_{\hat{\alpha}}) \nonumber
\end{align}
where we also used the constraint $\sum_K p^a_K=1$ $\forall a$.
Under the constraints $0\leq p^a_K\leq 1$ the maximization of $F$ yields
$p^+_{\hat{\alpha}}=p^-_{\hat{\alpha}}=1$ and
$F=(\sqrt{d_+}+\sqrt{d_-})^2/d^4$ in agreement with \cite{cloning}.

\subsection{$1\rightarrow N$ Cloning of $SU(2)$ transformations}\label{su2clon}
\begin{figure}[h]
      \begin{center}
    \includegraphics[width=8cm ]{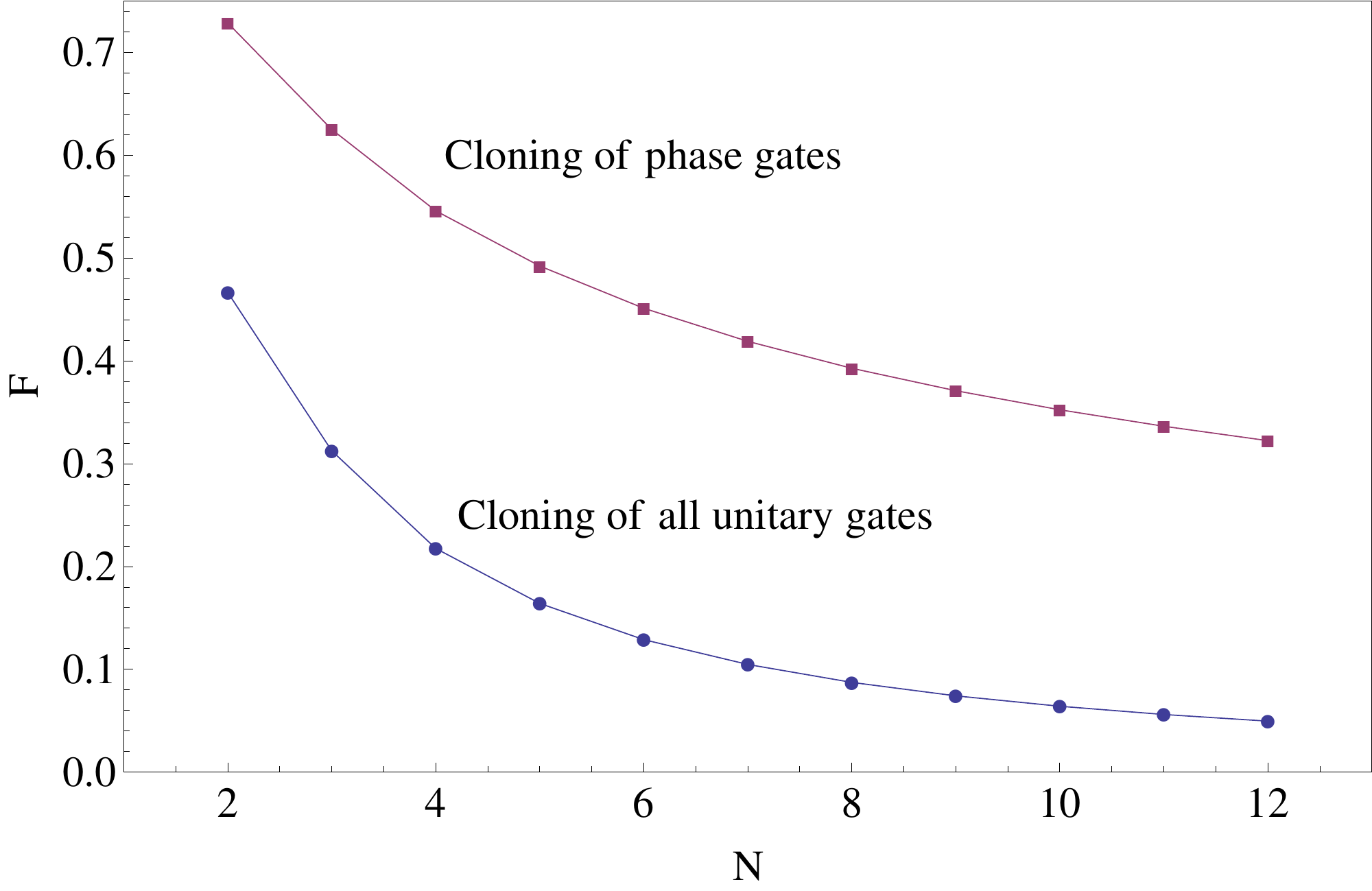}
    \caption{Average process fidelity of optimal $1\rightarrow N$
      cloning of a qubit channel.  The bottom line shows optimal
      process fidelity for cloning of all qubit unitary channels,
      whereas the top line corresponds to optimal cloning of only
      qubit phase gates.}
     \label{fig:clongraph}
  \end{center}
\end{figure}

Cloning of qubit unitary gates might be viewed as a transformation of
the defining representation $U$ of $SU(2)$ into the reducible
representation $U^{\otimes N}$. The representation $U^{\otimes N}$
decomposes into irreps as:
\begin{align}
\label{eq:su2decomp}
U^{\otimes N}=\bigoplus_{a=\bb N/2 \kk}^{N/2} U_a \otimes \one_{m_a},
\end{align}
where $\bb x \kk$ denotes the fractional part of $x$ (i.e. $\bb N/2
\kk$ is $0$ for $N$ even and $1/2$ for $N$ odd) and
$m_a=\frac{2a+1}{N/2+a+1} \binom{N}{N/2+a}$ \cite{su2decomp}.  Since
the input representation has $\beta=1/2$, the irreps in $U^{\otimes
  N}\otimes U^*$ are labelled by $K$ ranging from $\bb (N+1)/2 \kk$ to
$(N+1)/2$. In particular, each value of $K$ derives either from
$a=K-1/2$ or from $a=K+1/2$. The only exceptions to this rule are the
maximum $K$ and $K=0$ for odd $N$, which derive from a single
value of $a$.  This simplifies the problem and we can rewrite it as the
maximization of
\begin{align}
  F=&\sum_{K=\bb \frac{N}{2} \kk+\frac12}^{\frac{N-1}{2}} \left(\sqrt{q^{K-\frac12}_K x_K}+ \sqrt{q^{K+\frac12}_K (1-x_{K+1})}\right)^2 +\nonumber \\
  &+q^{\frac{N}{2}}_{\frac{N+1}{2}} x_{\frac{N+1}{2}}+2\bb \tfrac{N}{2}
  \kk q^{\frac12}_0 (1-x_0)
\end{align}
with respect to $0\leq x_K \leq 1$, where we denoted $x_K\equiv
p^{K-1/2}_{K}$ and consequently $p^{K+1/2}_{K}\equiv 1-x_{K+1}$ due to
the normalization constraints (\ref{eq:optprogfinal}).  Thus, for a
given $N$ we need to optimize roughly $N/2$ parameters $x_K$. This can
be done analytically by symbolic calculus for small values of $N$ or
numerically. In Fig.~\ref{fig:clongraph} the optimal fidelity is
plotted for $N$ up to $12$.

\subsection{$1\rightarrow N$ Cloning of qubit phase gates} \label{1toNclonphase}

The third application of the general method that we show here is
cloning of qubit phase gates, i.e.~unitary transformations
$U=diag(1,e^{i\phi})$ that are diagonal in the computational basis
$\{\ket{0},\ket{1}\}$. In this case the input representation $U$ of
$U(1)$ is reducible, and it is transformed into the different
reducible representation $U^{\otimes N}$.  Since $U(1)$ has only
$1$-dimensional irreps we have $d_a=d_\beta=d_K=1$.  We can decompose
$U^{\otimes N}$ as $U^{\otimes N}=\bigoplus_{a=0}^{N} e^{i a
  \phi}\pi_a$, where $\pi_a$ denotes the projection on the subspace
spanned by tensor products of vectors in the computational basis with
$a$ factors equal to $\ket1$.  Consequently, $U^{\otimes N}\otimes
U^*$ contains representations $e^{i K \phi}$ $K=-1,\ldots,N$ and each
$K\in \{0,\ldots,N-1\}$ can be obtained either from $(a,\beta)=(K,0)$
or from $(a,\beta)=(K-1,1)$. The irreps $K=-1$ and $K=N$ can derive
only from one pair $(a,\beta)$.  This allows us to rewrite the problem
as the maximization of the following expression
\begin{align}\label{eq:maxcphace}
F&=\sum_{K=0}^{N-1} \left(\binom{N}{K}\sqrt{x_K}+
  \binom{N}{K+1}\sqrt{1-x_{K+1}}\right)^2 + \nonumber \\
&  +(1-x_0)+x_{N}
\end{align}
with respect to $0\leq x_K \leq 1$, where we denoted $x_K\equiv
p^{K}_{K}$ and $p^{K+1}_{K}\equiv 1-x_{K+1}$ thanks to the
normalization constraints (\ref{eq:optprogfinal}).  We performed the
optimization for small values of $N$ by symbolic calculus. As one
could expect, the optimal fidelity for $1\rightarrow N$ cloning of
phase gates is much better than the one for cloning of arbitrary qubit
unitary channels, as illustrated in Fig.~\ref{fig:clongraph}.

\subsection{Realization of $1\rightarrow 2$ cloning of qubit phase gates}
In this section we discuss physical schemes for the realization of
optimal $1\rightarrow 2$ cloning of qubit phase gates. Before
describing our proposals, let us summarize the results implied by the
previous sections. In the case of $N=2$ the maximization of Eq.
(\ref{eq:maxcphace}) yields $x_0\equiv p^0_0=1, x_1\equiv p^1_1=1/2,
x_2\equiv p^2_2=0$, 
which using equation (\ref{eq:optimalcomb}) gives
\begin{align}
\label{eq:optr}
\begin{split}
  R&=\ket{\psi_0}\bra{\psi_0}+\ket{\psi_1}\bra{\psi_1}+ \\
  & + \left(\frac12 P^{1,0}_1+P^{0,0}_0\right)\otimes \frac13 (P^{0,1}_{-1}+ \frac12 P^{1,1}_0 + P^{2,1}_1)+ \\
  & + \left( P^{2,1}_1+\frac12 P^{1,1}_0\right)\otimes\frac13
  (P^{0,0}_0+ \frac12 P^{1,0}_1 + P^{2,0}_2),
\end{split}
\end{align}
where we defined
\begin{align*}
  \ket{\psi_0}&=\ket{000000}+ \frac{1}{\sqrt{2}}\ket{011101}+\frac{1}{\sqrt{2}}\ket{101110}\\
  \ket{\psi_1}&=\ket{111111}+
  \frac{1}{\sqrt{2}}\ket{010001}+\frac{1}{\sqrt{2}}\ket{100010},
\end{align*}
we used notation $\ket{a\overline{a}\beta\gamma
  d\overline{d}}\equiv \ket{a
  \overline{a}}_{0}\ket{\beta}_{1}\ket{\gamma}_{2}\ket{d\overline{d}}_{3}$
and the tensor products are ordered as $X\otimes Y \in
\mathcal{L}(\hilb{H}_0 \otimes \hilb{H}_1)\otimes
\mathcal{L}(\hilb{H}_2 \otimes \hilb{H}_3)$.  Let us
evaluate $R*|U\kk \bb U|$, which corresponds to an overall channel between
$\hilb{H}_0$ and $\hilb{H}_3$ that is created after the unitary gate
$U=diag(1,e^{i\phi})$ is inserted into the cloning circuit. All the
terms $P^{a,\beta}_{K}\otimes P^{d,\gamma}_L$ in Eq. (\ref{eq:optr})
do not contribute, since they have $\beta \neq \gamma$ and $|U\kk$
contains only the terms $\ket{\beta}_{1}\ket{\beta}_2$. Thus, we obtain
\begin{align}
&R *|U\kk \bb
U|=\ket{\varphi_0}\bra{\varphi_0}+\ket{\varphi_1}\bra{\varphi_1} \\
  \begin{split}
&\ket{\varphi_0}=\ket{0000}+\frac{e^{i\phi}}{\sqrt{2}}(\ket{0101}+\ket{1010})\\
&\ket{\varphi_1}=e^{i\phi}\ket{1111}+ \frac{1}{\sqrt{2}}(\ket{0101}+\ket{1010}). 
  \end{split} \nonumber
\end{align}
One can check by direct calculation that this channel is achieved by the following quantum circuit
\begin{align} 
  \begin{aligned}
    \Qcircuit @C=1em @R=1em {
      &\qw &\qw &\qw &\ctrl{2} &\ctrl{1} &\qw &\qw &\ctrl{2} & \qw &\qw \\
      &\qw &\qw &\ctrl{1} &\qw &\ctrl{1} &\qw &\ctrl{1} &\qw & \qw &\qw \\
      & &\lstick{\ket{0}} & \gate{H} & \gate{H} p&\targ &\gate{U(\phi)} &\targ &\targ & \measureD{\;} &
        }
  \end{aligned},
\end{align}
where $H$ denotes the Hadamard gate and the ancillary qubit is
prepared in state $\ket{0}$. The dimension of quantum system that is
used in parallel with the action of the cloned gate is called quantum
memory in the context of quantum protocols \cite{memorycost}. In the
proposed circuit the memory is four dimensional ($2$ qubits).  In
order to make the memory smaller, one can employ the techniques from
ref. \cite{memorycost} that are based on the covariance of the
problem. In this way one can trade a four dimensional quantum memory
for a three dimensional memory and one bit of classical
communication. We were able to describe such a memory efficient
realization of the optimal cloning of a phase gate in terms of
isometries (see Figure \ref{sch:finprob}),
\begin{align}\label{sch:finprob}
  \begin{aligned}
    \Qcircuit @C=0.6em @R=0.4em {
      &\qw&\multigate{2}{\;V\;}&\ustick{1}\qw & \qw &\gate{\mathcal{E}_g}& \ustick{2} \qw & \multigate{2}{\;Q_i\;} \qw& \qw & \qw & \qw\\
      &\ustick{0} \qw&\ghost{\;V\;}&\ustick{A}\qw & \qw              &\qw          &\qw              & \ghost{\;Q_i\;}&\ustick{3} \qw &\qw& \qw \\
      & &\pureghost{\;V\;}&\ustick{B} \qw & \measureD{\;} &\cw &\cw &
      \pureghost{\;Q_i\;} \cw &\ustick{C} \qw& \measureD{\;} & }
  \end{aligned}
\end{align}
however synthesizing a corresponding quantum circuit goes beyond the
scope of this manuscript. The isometry $V$ in Figure \ref{sch:finprob}
is the following
\begin{align}
  V&=\ket{0}_{B} \left(\ket{1}\ket{0}\bra{00} + \frac{1}{\sqrt{2}}\ket{2}\ket{1}\bra{01} + \frac{1}{\sqrt{2}}\ket{3}\ket{1}\bra{10}\right)+ \nonumber \\
  & +\ket{1}_{B} \left(\ket{1}\ket{1}\bra{11} +
    \frac{1}{\sqrt{2}}\ket{2}\ket{0}\bra{01} +
    \frac{1}{\sqrt{2}}\ket{3}\ket{0}\bra{10}\right) \nonumber
\end{align}
where the shortened expressions $\ket{1}\ket{0}\bra{00}$ stand for
$\ket{1}_{A}\ket{0}_{1}\bra{00}_{0}$ and the subsystems $A$, $B$ are a
qutrit, and a qubit, respectively.  The result of the measurement in
the $\{ \ket{0}, \ket{1} \}$ basis determines whether $Q_0$ or $Q_1$
will be used after the action of the input gate. The isometries $Q_0$
and $Q_1$ are defined as follows
\begin{align}
 Q_0&=\ket{1}_{C} \Big(\ket{00}\bra{0}\bra{1} + \ket{01}\bra{1}\bra{2} + \ket{10}\bra{1}\bra{3}\Big) +\nonumber \\
 & +\ket{2}_{C} \ket{11}\bra{1}\bra{1} + \ket{3}_{C} \ket{00}\bra{0}\bra{2} + \ket{4}_{C}\ket{00}\bra{0}\bra{3}\nonumber
\end{align}
\begin{align}
  Q_1&=\ket{1}_{C} \Big(\ket{11}\bra{1}\bra{1} + \ket{01}\bra{0}\bra{2} + \ket{10}\bra{0}\bra{3}\Big) +\nonumber \\
  & + \ket{2}_{C} \ket{00}\bra{0}\bra{1} +
  \ket{3}_C\ket{11}\bra{1}\bra{2} + \ket{4}_{C}
  \ket{11}\bra{1}\bra{3}, \nonumber
\end{align}
where we shortened $\ket{00}_3\bra{0}_2\bra{1}_A$ as
$\ket{00}\bra{0}\bra{1}$ and the ancillary quantum system $C$ is four
dimensional.

\section{Conclusions}
\label{sec:conclusions}

In this paper we reviewed the general theory of higher order quantum
maps and within this framework we addressed a general class of quantum
computational tasks involving the processing of unitary channels.  We
considered the scenario in which one has access to a single use of $N$
unknown unitary channels $\{ \mathcal{U}^{(i)}_g \}_{i=1 \dots N}$ in
an arbitrary sequential order, where the action of each
$\mathcal{U}^{(i)}_g$ on a state $\rho $ is described by a unitary
representation ${U}^{(i)}_g $ of a fixed compact group $G$, i.e.
$\mathcal{U}^{(i)}_g(\rho) = {U}^{(i)}_g \rho {U}^{(i) \dag}_g$.  The
task we considered is to exploit the uses of the unitary channels $\{
\mathcal{U}^{(i)}_g \}_{i=1 \dots N}$ to create a target unitary
channel $\mathcal{V}_g $ which is described by a different unitary
representation $V_g$ of the same group $G$.  As a figure of merit we
chose the group average of the channel fidelity between the output
channel and the ideal one. We proved that the optimal scheme does not
require any non-circuital higher order map, but it can be realized by a
three-step protocol: i) application of a preprocessing channel
$\mathcal{C}_1$, ii) parallel application of the unitary channels
$\mathcal{U}^{(i)}_g$ and iii) final application of a postprocessing
channel $\mathcal{C}_2$.

Moreover, we rephrased the circuit optimization problem as simplified
semidefinite programming that significantly reduces the number of
variables involved in the optimization, as can be appreciated by
comparing the original formulation of the problem in
Eq.~\eqref{eq:optprob1} and the simplified one in
Eq.~\eqref{eq:optprogfinal}. One can see, for example, that in the
case of $1 \to N$ cloning of a $SU(2)$ gate (see Section
\ref{su2clon}) the number of parameters $D$ in the semidefinite
program exponentially reduces from $D\sim 2^{2N}$ to $D\sim N^2$.
Remarkably, the results of Proposition \ref{prop:optimality} along
with the results of Ref.~ \cite{memorycost} allow us to assess an
upper bound to the amount of quantum memory which must be kept
coherent from the optimal preprocessing to the postprocessing phase
through the parameter $\max_K m_K$, the maximal multiplicity in the
decomposition of Eq.~\eqref{eq:decompohilbspace}.

The quantum processing task that we consider in this paper is very
general and includes a number of interesting scenarios as special
cases. Indeed, in section \ref{sec:examples}, besides recovering
the results of $1 \to 2$ cloning of $SU(d)$ unitaries, we provided the
optimal solution for the task of transforming an $SU(2)$ irrep into a
different one, and for the $1 \to N$ cloning of $SU(2)$ and $SU(1)$.
The last two cases illustrate how a stronger prior knowledge about
the unknown unitaries enables a higher fidelity (see Fig. \ref{fig:clongraph}) in the same way as it
happens for phase covariant \cite{phaseclon} versus universal state
cloning \cite{cloningbuzek,cloningwerner}.

An alternative way to achieve the transformation from
 $\{ \mathcal{U}^{(i)}_g \}_{i=1 \dots N}$ to $\mathcal{V}_g $ is to
estimate $g$ and then to prepare the estimated unitary.  This
measure and prepare strategy can be generally more easily implemented than the
pre and post-processing one and has the advantage that it could be
applied even in the case in which the uses $\{ \mathcal{U}^{(i)}_g
\}_{i=1 \dots N}$ and the quantum state $\rho$ which $\mathcal{V}_g $
will be applied to, are not available at the same time. Because of
that, there can be situations in which one could prefer to apply the
measure-and-prepare strategy if the consequent performance loss is below a
given threshold. Whithin this perspective it would be useful to
characterize under which conditions this two strategies achieve similar
fidelity.
Especially interesting would be the study of  the asymptotic scaling
of the optimal $N \to M$ cloning of unitaries and to verify whether
the two startegies exhibit the same scaling for $M \to \infty$.
This would be a generalization of the known result of the asymptotic
convergence of optimal state cloning to state estimation .
The results of the current paper provide versatile tools for the study
of this problem and this investigation will be the subject of future works.

\appendix

\section{Irrelevance of the multiplicity spaces}\label{a:multiplicity}

Our aim is to show that two sets of channels $\{\mathcal{U}_g: g\in
G\}, \{\mathcal{U}'_g:g\in G\}$ defined by two representations of
group $G$ that differ only in the multiplicities of the irrep's are
mutually perfectly transformable. This statement is made precise in
the following lemma.

\begin{lemma}
\label{th:convertibility}
A set of unitary channels $\mathcal{U}_g$ defined by a representation
$U_g=\bigoplus_{\beta} U^{[\beta]}_g \otimes \one_{m_\beta}$ and a set
of unitary channels $\mathcal{U}'_g$ defined by a representation
$U'_g=\bigoplus_{\beta} U^{[\beta]}_g$ are perfectly mutually
transformable, i.e. there exist two deterministic $2$-combs $R$ and $\widetilde{R}$ such
that
\begin{align}
|U'_g \kk \bb U'_g|&= R* |U_g \kk \bb U_g| \quad \forall g\in G \nonumber \\
|U_g \kk \bb U_g|&= \widetilde{R}* |U'_g \kk \bb U'_g| \quad \forall g\in G.
\end{align}
\end{lemma}

\Proof The proof is constructive. For the construction of $R$ we
define two channels $\mathcal{X},\mathcal{Y}$ such that
\begin{align}\label{sch:etof}
  \begin{aligned}
    \Qcircuit @C=0.6em @R=0.4em {
      &\gate{\mathcal{U}'_g}& \qw }
  \end{aligned}
\; = \;
 \begin{aligned}
    \Qcircuit @C=0.6em @R=0.4em {
      &\gate{\mathcal{X}}&\qw&\qw&\gate{\mathcal{U}_g}& \qw & \gate{\mathcal{Y}} &\qw }
  \end{aligned}.
\end{align}
The channel $\mathcal{X}$ is an isometry that embeds the Hilbert space
$\bigoplus_{\beta} \hilb{H}_{\beta}$ in a subspace $\pi$ of
$\bigoplus_{\beta} \hilb{H}_{\beta} \otimes \mathbb{C}^{m_\beta}$
defined by the choice of a single vector $\ket{\phi_\beta}$ in each
multiplicity space. The channel $\mathcal{Y}$ has Kraus operators the
inverse isometry
$V^\dag:=\sum_\beta(I_\beta\otimes\bra{\phi_\beta})\Pi_\beta$ and
$K_i\sqrt{I-VV^\dag}$ where $\Pi_\beta$ represents the projection on
the subspace $\hilb{H}_\beta\otimes\mathbb{C}^{m_\beta}$, while $K_i$
are Kraus operators of an arbitrary trace-preserving map from the
support of $I-VV^\dag$ to $\bigoplus_{\beta} \hilb{H}_{\beta}$.
Finally, $R:=X\otimes Y$, with $X,Y$ being Choi matrices of
$\mathcal{X}$,$\mathcal{Y}$.

For the construction of $\widetilde{R}$ we define an ancillary system
$M$ and two channels $\widetilde{\mathcal{X}},\widetilde{\mathcal{Y}}$
as follows
\begin{align}\label{sch:ftoe}
  \begin{aligned}
    \Qcircuit @C=0.6em @R=0.4em {
      &\gate{\mathcal{U}_g}& \qw }
  \end{aligned}
\; = \;
 \begin{aligned}
    \Qcircuit @C=0.6em @R=0.4em {
      &\multigate{1}{\;\widetilde{\mathcal{X}}\;}&\qw&\qw&\gate{\mathcal{U}'_g}& \qw & \multigate{1}{\;\widetilde{\mathcal{Y}}\;} &\qw\\
      &\pureghost{\;\widetilde{\mathcal{X}}\;}&\ustick{M} \qw&\qw             &\qw          &\qw              & \ghost{\;\widetilde{\mathcal{Y}}\;} }
  \end{aligned}.
\end{align}

We set the dimension of $\hilb{H}_M$ to be $\max_{\beta} m_\beta$. The
channel $\tilde {\mathcal X}$ is just an isometric embedding of
$\bigoplus_{\beta} \hilb{H}_{\beta}\otimes\mathbb C^{m_\beta}$ into
$\bigoplus_{\beta} \hilb{H}_{\beta} \otimes \mathbb{C}^M$. The channel
$\tilde{\mathcal Y}$ is now analogous to $\mathcal Y$, with the only
difference that one projects $\hilb{H}_\beta\otimes\mathbb C^M$ into
$\hilb{H}_\beta\otimes \mathbb C^{m_\beta}$ and its orthogonal
complement. \qed

\acknowledgments MS was supported by the COST Action MP1006,
APVV-0646-10 (COQI), VEGA 2/0127/11 (TEQUDE) and by the Operational
Program Education for Competitiveness - European Social Fund (projects
No.  CZ.1.07/2.3.00/30.0004 and CZ.1.07/2.3.00/20.0060) of the
Ministry of Education, Youth and Sports of the Czech Republic.


\begin{thebibliography}{99}
\bibitem{compact}
The hypothesis that $G$ is a compact group guarantees the existence of
an invariant Haar measure $\d g$.
\bibitem{subspacedim}
The Hilbert space $m_N$ has the same dimension as the support of the operator $R^{(N)}$.
\end{thebibliography}
\end{document}